\documentclass[fleqn,usenatbib,useAMS]{mnras}

\usepackage{graphicx}	% Including figure files
\usepackage{amsmath}	% Advanced maths commands
\usepackage{amssymb}	% Extra maths symbols
\usepackage{multicol}        % Multi-column entries in tables
\usepackage{bm}		% Bold maths symbols, including upright Greek
\usepackage{pdflscape}	% Landscape pages

\usepackage[T1]{fontenc}
\usepackage{ae,aecompl}

\usepackage{newtxtext,newtxmath}
\title[Blanco DECam Bulge Survey ]{The Blanco DECam Bulge Survey. I. The Survey Description and Early Results}
\author[R. M. Rich et al.]{
R. Michael Rich,$^{1}$\thanks{E-mail: rmr@astro.ucla.edu (RMR)}
Christian I. Johnson,$^{2}$
Michael Young,$^{3}$
Iulia T. Simion,$^{4}$
\newauthor
William I. Clarkson,$^{5}$
Catherine Pilachowski, $^{6}$
Scott Michael,$^{3}$
Andrea Kunder,$^{7}$
\newauthor
A. Katherina Vivas,$^{8}$
Andreas Koch,$^{9}$
Tommaso Marchetti,$^{10}$
Rodrigo Ibata,$^{11}$
\newauthor
Nicolas Martin,$^{11}$
Annie C. Robin,$^{12}$
Nad\`ege Lagarde,$^{12}$
Michelle Collins,$^{13}$
\v{Z}eljko Ivezi\'{c},$^{14}$
\newauthor
Roberto de Propris,$^{15}$
Juntai Shen,$^{16}$
Ortwin Gerhard,$^{17}$
and Mario Soto,$^{18}$
\\
$^{1}$Department of Physics \& Astronomy, Univ. of California Los Angeles, PAB 430 Portola Plaza, Los Angeles, CA 90095-1547, USA\\
$^{2}$ Space Telescope Science Institute, 3700 San Martin Drive, Baltimore, MD 21218 \\
$^{3}$Indiana University, University Information Technology Services, CIB 2709 E. 10th Street, Bloomington, IN 47401 USA \\
$^{4}$ Key Laboratory for Research in Galaxies and Cosmology,Shanghai Astronomical Observatory, 80 Nandan Road, Shanghai 20030, China \\
$^{5}$ Department of Natural Sciences, University of Michigan-Dearborn, 4901 Evergree Rd, Dearborn, MI 48128 \\
$^{6}$ Indiana University Department of Astronomy, SW319, 727 E 3rd Street, Bloomington, IN 47405 USA \\
$^{7}$ College of Arts \& Sciences, St Martin's University, Ernsdorff Center 130
5000 Abbey Way SE Lacey, WA 98503\\
$^{8}$ Cerro Tololo Inter-American Observatory, NSF's National Optical-Infrared Astronomy Research Laboratory, Casilla 603, La Serena, Chile\\
$^{9}$Zentrum f\"ur Astronomie der Universit\"at Heidelberg, Astronomisches Rechen-Institut, 69120 Heidelberg, Germany\\
$^{10}$ European Southern Observatory, Karl Schwarzschild-Strasse 2, 85748 Garching bei Munchen, Germany\\
$^{11}$Observatoire de Strasbourg,11 rue de l'observatoire, 67000 Strasbourg, France\\
$^{12}$Institut Utinam, CNRS UMR 6213, Univ. Bourgogne Franche-Comt\'e, OSU THETA, Observatoire de Besancon, BP 1615 25010 Besancon Cedex, France \\
$^{13}$Department of Physics, Univ. of Surrey, Surrey GU2 7XH, United Kingdom\\
$^{14}$University of Washington, Astronomy Dept, Box 351560, 3910 15th Av. NE, Seattle WA 98195-1580, USA\\
$^{15}$Finnish Center for Astronomy with ESO, Turku University, Vesilinnantie 5, Turku, Finland\\
$^{16}$Department of Astronomy, School of Physics and Astronomy, Shanghai
Jiao Tong University, 800 Dongchuan Road, Shanghai 200240, China and \\
Shanghai Key Laboratory for Particle Physics and Cosmology, 200240,
Shanghai, China  \\
$^{17}$ Max-Planck-Institut f\"ur Extraterrestrische Physik, Giessenbachstrasse, D-85748 Garching, Germany \\
$^{18}$ Instituto de Astronom'a y Ciencias Planetarias, Universidad de Atacama, Copayapu 485, Copiap\'o, Chile}

\date{Accepted XXX. Received YYY; in original form ZZZ}

\pubyear{2020}

\begin{document}
\label{firstpage}
\pagerange{\pageref{firstpage}--\pageref{lastpage}}
\maketitle

% Abstract of the paper
\begin{abstract}
The Blanco Dark Energy Camera (DECam) Bulge survey is a Vera Rubin Observatory (LSST) pathfinder imaging survey, spanning $\sim 200$ sq. deg. of the Southern Galactic bulge,  $-2^\circ <$b$< -13^\circ$ and $-11^\circ <$l$ < +11^\circ$.  We have employed the CTIO-4m telescope and the Dark Energy Camera (DECam) to image a contiguous $\sim 200$ sq. deg. region of the relatively less reddened Southern Galactic bulge, in SDSS $u$ + Pan-STARRS$grizy$.    Optical photometry with its large colour baseline will be used to investigate the age and metallicity distributions of the major structures of the bulge.  Included in the survey footprint are 26 globular clusters imaged in all passbands.  Over much of the bulge, we have Gaia DR2 matching astrometry to $i\sim 18$, deep enough to reach the faint end of the red clump.  This paper provides the background, scientific case, and description of the survey.  We present an array of new reddening-corrected colour-magnitude diagrams that span the extent of Southern Galactic bulge.  We argue that a population of massive stars in the blue loop evolutionary phase, proposed to lie in the bulge, are instead at $\sim 2$ kpc from the Sun and likely red clump giants in the old disk.  A bright red clump near $(l,b)=(+8^\circ,-4^\circ)$ may be a feature in the foreground disk, or related to the long bar reported in earlier work.    We also report the first map of the blue horizontal branch population spanning the BDBS field of regard, and our data does not confirm the reality of a number of proposed globular clusters in the bulge.

\end{abstract}

% Select between one and six entries from the list of approved keywords.
% Don't make up new ones.

%%%%%%%%%%%%%%%%%%%%%%%%%%%%%%%%%%%%%%%%%%%%%%%%%%

%%%%%%%%%%%%%%%%% BODY OF PAPER %%%%%%%%%%%%%%%%%%

% The MNRAS class isn't designed to include a table of contents, but for this document one is useful.
% I therefore have to do some kludging to make it work without masses of blank space.
%\begingroup
%\let\clearpage\relax
%\tableofcontents
%\endgroup
%\newpage

\section{Introduction}
\subsection{Survey Motivation}
The goal of BDBS is to use the Dark Energy Camera (DECam; \citet{flaugher15}) to produce an optical multiband map of the Galactic bulge South of the Galactic plane, where extinction is significantly lower and more uniform than in the North.  The Northern bulge still can be studied and our team is proposing to do so.  However, the lower extinction of the Southern bulge has meant that nearly all spectroscopic and imaging surveys have focused on that region.   Hence this pathfinder survey emphasizes the Southern bulge.  

The VVV survey has mapped the bulge and disk in the near-infrared, however, BDBS uses optical colours including the $u$ and $Y$ bands.  These choices result in a uniformly calibrated optical dataset obtained with a single instrument.  Further, the colour baseline enables far superior metallicity estimates, compared with infrared surveys.  The ultimate science aim is to also exploit the Gaia database and other approaches 
to separate the foreground disk population from the bulge.   In \citet{bdbs2} we report metallicities for red clump giants from the $(u-i)_0$ colour and we ultimately aim to statistically subtract the foreground disk population so that the bulge main sequence turnoff (and population age) can be constrained over the full field.  Hence there is the real possibility of mapping the metallicity and age as a function of position in the bulge, including the X-structure.   This will provide important constraints on the formation history of the bulge.

Galaxy bulges and spheroids account for most of the light from luminous galaxies, and any understanding of galaxy evolution must include comprehension of their formation process \citep[e.g.][]{rich2013, mcwilliam16, barbuy18}. The bulges hosted by disk galaxies appear to be divided into spheroidal bulges and those that are barred or disk-originated or "secular" \citet{kormendy04}; the evidence strongly suggests our bulge is disk-originated (see e.g. \citet{shen10}). 
In principle, the Galactic bulge should be the calibrator for studying bulges across cosmic time, but the high extinction, extreme crowding, superposition of stellar populations along the line of sight, and an intrinsically complex environment, render the bulge observationally complicated.
Since 2000, observations have conclusively shown that the majority of mass in the bulge is in the bar e.g. \citep{shen10}, but many problems in the formation and chemical evolution of the bulge remain both unsolved and debated.  Nonetheless, the Milky Way bulge is $\sim 100$ times closer than the nearest bulge hosted in a comparable galaxy, M31.  Consequently it is possible to measure proper motions, radial velocities, compositions,
and eventually parallaxes for stars spanning the HR diagram from dwarfs to giants; such observations are unlikely to be possible for decades, in the bulge of M31.

The grasp of the the Dark Energy Camera represents a dramatic advance in imaging area over prior facilities, and is within striking distance of the LSST capabilities.  At the time the survey was conceived, the authors appreciated that many significant problems in the bulge required uniformly calibrated photometry spanning a wide area.  Further, such photometry should include the UV/optical colours that were not observed in the VVV study. Optical colours would offer the prospect of a wide colour baseline that is better applicable to the derivation of photometric metallicities.  As was shown in the early work of \citet{fwr84}, the infrared $J-K$ colour has an insufficient baseline to yield  good enough constraints on metallicities to make meaningful conclusions about the distribution function.  Our aim is to use the photometry of millions of stars to replace the coarse metallicity map derived from $J-K$ photometry \citep{gonzalez12} with maps employing the wider colour baseline of an optically derived metallicity distribution function, spanning as much of the bulge/bar as possible.  

Several specific, significant developments motivated this survey.  First, the discovery of an X-shaped structure in the Southern Galactic bulge \citep{mcwzoc10, nataf10} revealed especially clearly in the analysis of \citet{wegg13}.   Widespread, yet widely varying claims for multimodality in the bulge abundance distribution is a second driver.   Multimodality was first reported by \citet{vhill11}, and claims for multimodality in the distribution of [Fe/H], with consequent interpretation, have blossomed (see e.g. Table 2  and Figure 4 in \citet{barbuy18}.  The most extreme claim for multimodality is presented in \citet{bensby17}, which argues for five distinct peaks in the abundance distribution.    One might reasonably expect such an extended and complicated star formation history to be problematic, even in terms of the relatively limited observations to date.   If due to multiple starbursts, one might reasonably expect to find the more metal rich populations to be strongly enhanced in s-process elements that would be produced in multiple generations of AGB stars; no such enhancements are observed in the bulge \citep{johnson14,mcwilliam16}.  \citet{ness13} also argues for 5 peaks of varying amplitude in the bulge metallicity distribution, with more metal rich stars concentrated toward the plane.  Their analysis presents a deconvolution of the metallicity distribution into multiple Gaussian peaks that attempts to describe components of the bulge metallicity distribution that vary with varying amplitudes as a function of distance from the plane.

There is also some significant range of opinion on how populations in the bulge are spatially distributed.  In addition to \citet{ness13}, \citet{zoc17} finds a sharp demarcation in the spatial distribution at [Fe/H]$\sim0$; more metal rich stars are confined strongly to the plane, while stars with [Fe/H]$<0$ are found to follow a spheroidal distribution.  The physical cause for such a dramatic change in spatial distribution at the somewhat arbitrary point of Solar metallicity is not immediately obvious.
On smaller scales, \citet{schultheis19} argue that the central 50 pc experiences a spike in the fraction of stars with [Fe/H]$>0$.   One of the key science goals of BDBS is to apply the photometric metallicities derived for millions of stars to test these and similar conclusions, which have been drawn from a range of spectroscopic surveys that have relatively modest numbers of stars with derived abundances.  Further, by spanning a large area, we are able to include the lower extinction areas of the bulge and to match significant numbers of our stars to the Gaia catalog.  As described here and in \citet{bdbs2}, BDBS also attains the main sequence turnoff.  With future proper motion cleaning or foreground population subtraction, we may also be able to use the main sequence turnoff to map the age of the bulge as well.

The age of the bulge has been controversial for decades.  Early constraints came from the presence of RR Lyrae stars \citep{baade51} linking at least part of the bulge population to old, metal poor globular clusters.  The early discovery of large numbers of M giants in the direction of the Galactic Center as described in \citet{blanco65} led to a prescient classification of the bulge at the landmark 1957 Vatican symposium on Stellar Populations as "old disk" population.  As described earlier, Victor and Betty Blanco used the capabilities of the CTIO 4m telescope to study the Mira variable population in landmark early bulge studies.  The earliest CCD photometry from the 4m telescope was obtained as the results of proposals by the author in the early 1980s; \citet{trw84,t88} revealed an apparently old main sequence turnoff in the Plaut $-8^\circ$ field, but were significantly confused by the foreground disk population; a plausible old main sequence turnoff was detected only at the higher latitude $-8^\circ$ and $-10^\circ$ fields where foreground contamination is an issue.   The problem of the serious overlap of the foreground disk and the bulge main sequence turnoff was finally solved by statistically subtracting the disk population.  \citet{ort95} used this approach along with HST photometry and direct comparison with the luminosity function of the globular cluster NGC 6553 to make a convincing case for an old bulge population.  \citet{clarkson08}, following \citet{kr02}, subsequently used proper motion separation to reveal clearly an old main sequence turnoff in a low latitude bulge field.  \citet{renzini18} employed the same method to show that the luminosity functions of both metal poor and metal rich stars in the bulge are most consistent with a 10 Gyr old population.  Analysis of the VVV survey data supports further an old Galactic bulge \citep{surot19}, yet the age problem remains, with significant tension between the main sequence turnoff measurements and derived ages for microlensed dwarfs. 

While the cumulative evidence from studies of the luminosity function and proper motion cleaned main sequence turnoff are compelling, analysis of microlensed bulge dwarfs appears to tell a different story.  \citet{bensby17} reports photometric ages for 99 microlensed dwarfs in the bulge, and $\sim 50\%$ of those above Solar metallicity are found to be younger than 8 Gyr based on a self consistent spectroscopic analysis from high resolution spectra.  Such a significant population of intermediate age, metal rich stars might be expected to spawn significant numbers of luminous evolved stars.   The bulge hosts a large population of Mira variables with period > 400 days; these are attributed to an intermediate age population \citep{catchpole16}.   However, the demarcation between old and intermediate age populations at the 400 day period mark is not clearly derived from any definitive theoretical underpinning or unequivocally accepted observation and inferences on population
age might be metallicity dependent.
%add new reference on miras

Although the optical colours we use in BDBS are more sensitive to reddening, they are also more sensitive to age and metallicity; consequently the wide field optical survey of BDBS is an excellent complement to the pencil beam surveys of HST.  We believe that the metallicity sensitivity may give us a different perspective on the overall age problem, compared to the infrared age derivation based on  VVV \citep{valenti19}.
%(check VVV for age paper).

BDBS is not the first wide field optical survey of the bulge.  In addition to the wide field surveys of Pan-STARRS \citep{chambers16} and DECaps \citep{schlafly18}, there has been the pioneering work of \citet{saha19} in Baade's Window. However, we emphasize that the power of our study is that it covers the full range of optical passbands and spans essentially the range of galactic latitude and longitude comprising the Southern bulge, hence including much of the bulge's mass (see e.g. \citet{launhardt02}).   Our field of regard includes the complete footprints of many major spectroscopic surveys, beginning with the Bulge Radial Velocity Assay \citet{rich07,kunder12} and ARGOS (\citet{ness13}; Figures 1-3).   Further, we have taken pains to calibrate the full range of the stellar population from the lower main sequence to red giant branch tip, and we employ psf-fitting in the photometry.  The details of these advances are given in the companion paper \citep{bdbs2}.

\subsection{Early Survey Science Papers}

Our team is pursuing a set of early survey science that is based on the dataset as it currently stands, without complete reductions for the Sgr Dwarf or off-axis disk fields.  We have calibration images for the Sgr Dwarf spheroidal, and the off-axis disk fields will be used to statistically correct for contamination by the foreground disk population, important for studies of the age of the bulge. 

Paper I (this paper) gives an overview of the survey and early science
results, while Paper II \citep{bdbs2} reports on a new method to derive metallicity from red clump giants using the $(u-i)_0$ colour, and illustrates the use of the BDBS dataset to explore globular clusters.  The candidate dwarf galaxy FSR1758 is shown to be globular cluster.     Paper III (Johnson et al. 2020 in prep.) will address the spatial structure of the red clump giant population as a function of metallicity.   The BDBS collaboration is engaged in preparing papers on the 26 globular clusters in the BDBS footprint (presented below in Table 2) and kinematics, based on the crossmatch between BDBS and Gaia DR2.    Our goal is to undertake a public data release sometime in 2021.

\subsection{Historical Background}

After the commissioning  of the Cerro Tololo 4m telescope (now the Blanco Telescope), Victor and Betty Blanco carried out a series of pioneering surveys of the Galactic bulge and Magellanic clouds using photographic plates at the prime focus.  The advent of red-sensitive Kodak IV-N plates allowed for the first time an investigation of the evolved red giant branch using access to wide field, near-infrared imaging.  Further, the prime focus allowed very low resolution slitless spectrosopy (grism)  to record low resolution spectra of all of the late-type stars in a 45 arcmin diameter field.  These observations revealed a bulge dominated by late M giants and devoid of carbon stars; the latter were discovered in abundance throughout the lower metallicity, intermediate age Magellanic clouds \citep{bbm78}.  

The Johnson $R, I$ photometry from the CTIO 4m Kodak IV-N photographic plates provided the input catalog for subsequent spectroscopic investigations that founded the new era of bulge research \citep{whitfordrich83,rich88}.  The precise photometry and spectral classifcations by  \citet{bmb84} also made possible a flood of successor Galactic bulge and Magellanic cloud science; the late type stars discovered and identified would fuel many other significant infrared photometric investigations, the most notable (for the bulge) was the infrared study of the bulge red giant branch, \citet{fw87}.  

Betty Blanco undertook the first survey of Galactic bulge RR Lyrae stars since that of \citet{baade51}, also using 4m CTIO prime focus photographic plates \citep{bb84, bb92, bb97}.   Because the Galactic bulge at $\delta =-30^\circ$ transits at CTIO and is therefore observable for 8 hr in a night, Blanco's analysis was not affected by the aliasing that plagued Baade's observations from Mt. Wilson, so her study remained the best study of bulge RR Lyraes until the era of large microlensing surveys.  Her work produced a modern catalog of RR Lyrae variables in bulge fields at $-4^\circ$ and $-6^\circ$.  It is worth recalling that the magnitudes were measured using an iris photometer, one star and epoch at a time; positions measured painstakingly with a measuring engine and then plate solved.   

The Blanco DECam Bulge Survey memorializes and honors the visionary scientific contributions of Victor and Betty Blanco, and for Victor Blanco's role as Director of the Cerro Tololo Interamerican Observatory.   It is noteworthy that the very survival of CTIO depended not only on the successful construction and commissioning of the 4m and other telescopes, but also the delicate navigation of CTIO as an institution through turbulent political times \citep{blanco2001}.  The work of Victor and Betty Blanco also opened a new era of stellar populations studies, especially in the Galactic bulge.  Many of the significant advances in Galactic bulge research from 1988 to the 1990s were enabled by the precise spectral classifications and carefully measured astrometric positions for the bulge M giants undertaken by the Blancos.   Although \citet{arp65} had produced the first colour-magnitude diagram (CMD) of the field toward  NGC 6522 at $(l,b)=(1,-4)$; the combination of photometric uncertainties, reddening, and variation in metallicity rendered the CMD very difficult to interpret.   Blanco's unpublished $R,I$ photometric colour-magnitude diagram and resulting red giant branch that made possible selection of spectroscopic targets, and the subsequent advances based on that early work.

We now stand at the threshold of a new era of very large spectroscopic and photometric surveys.  Some, such as APOGEE \citep{apogee17} and VVV \citep{vvv10}, are complete or in advanced stages.  The Rubin Observatory (LSST; \citet{LSST19}) will see its first light in a few years.  The VLT MOONS infrared surveys \citep{moons14} and 4MOST \citep{4most16} optical surveys will also commence in the next few years. 

Examining the vast range of work ongoing or contemplated in the Galactic bulge, we realized that an optically calibrated, deep, wide field, point spread function-fit photometric dataset does not exist for the bulge.  BDBS provides that survey and is also a pathfinder for science that may eventually be feasible in the area of Milky Way science, using LSST.

\section{Observations and Survey Strategy}

Most of our observations were executed in the course  of CTIO programs under PI R. M. Rich.   2013A-580 (2013 June 1-6 and July 14-16),
2014A-529 (2014 July 14-22).  As described in \citet{bdbs2}, the only observing run under photometric conditions was 2013 June 1-6.  However, thanks to the survey design in which all field edges overlap, we are able to calibrate the full dataset.  Table 1 (online) gives a detailed journal of observations.  In total, we have $\sim$ 250M sources in $g$ and $r$.  We have matched BDBS with Gaia DR2, yielding 74,791,629 matches with a mean value of $\sim 0.12$ arcsec and 
$\sigma=0.15$ arcsec.   The smaller number of Gaia matches is due to the limiting magnitude of Gaia DR2, and the high source density in the Galactic bulge.  

\begin{table*}
\caption{Journal of Observations$^a$}
	\centering
	\label{tab:table1}
    \tabcolsep=0.2cm
    \footnotesize
\begin{tabular}{lccccl}
\hline
Date & RA (2000)  & Dec (2000) & Exp (sec) & mag & dataset ID  \\
   
\hline

2013-06-01	&	279.308417	&	-35.712611	&	75.0	&	i	&	c4d\_130601\_045903\_ooi\_i\_v2 \\ 
2013-06-01	&	272.349083	&	-31.435111	&	100.0	&	i	&	c4d\_130601\_042434\_ooi\_i\_v2 \\ 
2013-06-01	&	279.240875	&	-35.644000	&	75.0	&	r	&	c4d\_130601\_045500\_ooi\_r\_v2 \\
2013-06-01	&	279.074750	&	-33.984028	&	75.0	&	r	&	c4d\_130601\_054112\_ooi\_r\_v2 \\
2013-06-01	&	279.238208	&	-35.715806	&	1.0	&	Y	&	c4d\_130601\_051921\_ooi\_Y\_v2 \\
2013-06-01	&	270.887000	&	-30.036556	&	100.0	&	i	&	c4d\_130601\_071751\_ooi\_i\_v2 \\
2013-06-01	&	272.348833	&	-31.436167	&	100.0	&	r	&	c4d\_130601\_042224\_ooi\_r\_v2 \\
2013-06-01	&	130.752833	&	-0.006333	&	20.0	&	Y	&	c4d\_130601\_230540\_ooi\_Y\_v2 \\
2013-06-01	&	279.237458	&	-35.712500	&	75.0	&	z	&	c4d\_130601\_050448\_ooi\_z\_v2 \\
2013-06-01	&	270.891500	&	-30.023250	&	30.0	&	r	&	c4d\_130601\_040406\_ooi\_r\_v2 \\
\hline
\multicolumn{3}{l}{$^a$ Abridged; full table is in electronic form.}\\
\end{tabular}
\end{table*}

Our exposure times reflect a desire to reach a roughly uniform depth in all passbands, and usually were $3\times 75$ sec.  However, the less sensitive $u$ band received deeper imaging with exposure times of 150 sec.  For much of the field, we also acquired special sets of $6\times30$ sec exposures arranged in 30" radius hexagons.  These will be used to push the astrometric solution fainter than the limit of Gaia DR2 matching, which is currently limited to $i\sim 18$ mag.  At present, much of our field has a 7 year baseline.  A series of short exposures were acquired for all of the fields, some as short as 0.75 sec.  These enable us to calibrate the brightest red giants in each field.  A detailed discussion of exposure time choices and the calibration strategy is given in the companion paper \citet{bdbs2}.

We chose to invest extra effort in calibrating the DECam Y to the Pan-STARRS $y$ and DECam u to the SDSS $u$ bands; we accomplished this by imaging the SDSS Stripe 82 regions under photometric conditions and covering a range in $sec z$.  The $y$ band is useful because it is our reddest band at nearly 1 $\mu \rm m$, so it is an excellent "bridge" to the infrared as we plan to match with the GLIMPSE, WISE, and MSX datasets.   Photometry in the ultraviolet has a long history in its use as a constraint on stellar abundance- see e.g. \citet{els62}.  More recently, \citet{ivezic08} took advantage of the $u$ band in deriving metallicities of disk stars.  In our case, the $u$ band has proven valuable for metallicity derivation (see \citet{bdbs2} Paper II), and we are optimistic that it will prove useful in constraining age and stellar populations.  The blue band photometry will be useful for matching our dataset to GALEX and Chandra imagery.  

The reduction of the DECam images employed a modified version of DAOPHOT and at this point in the project, we have not done artificial star tests.    To do such would vastly increase the computational and bookkeeping scale of the project beyond what is feasible at this time.  

\subsection{Calibration and Reddening correction}

Although a complete, detailed, description of our photometric calibration and reddening corrections is given in Sections 2 and 3 of \citet{bdbs2}, we provide an overview here.  Several factors made the final photometric calibration of this dataset especially challenging.  Only a subset of data were obtained in photometric conditions, hence the absolute calibration had to be transferred from the fields with absolute calibration to those fields imaged under non-photometric conditions; this was a challenging procedure due to the large numbers of individual images in our campaign.

When conditions were photometric, we obtained images of the Stripe 82 SDSS calibration field at a range of airmass and at regular intervals,
during photometric nights. Atmospheric extinction corrections were applied to all calibration
and science instrumental magnitudes using the airmass values
provided in the image headers and the extinction coefficients
provided by NOAO. For the $ugriz$ filters, the Stripe
82 instrumental magnitudes were matched to the calibrated
AB photometry from SDSS \citep{alam15} while the $Y$-band
data were matched to the UKIRT Infrared Deep Sky
Survey Large Area Survey (UKIDSS-LAS; \citet{lawrence07}). 

\subsubsection{Reddening}

It is well known that the reddening toward the bulge is high and spatially complex; we address this problem by adopting the high resolution $1^\prime\times1^\prime$ reddening map derived in Section 3 of  \citet{simion17} from the VVV survey, following the lower resolution sampling approach used in \citet{gonzalez11}.
The red clump giants in the
VVV are used to build a reddening map of higher sampling than that of the original survey,
by comparing the mean colour of the
red clump (RC) stars observed in each field with that of the RC population
in Baade's window $(l,b)=(0.9^\circ,-3.9^\circ)$.  While this approach provides a relatively high spatial resolution reddening correction, it does so only over the footprint of the VVV survey, $\lvert l \rvert < 10^\circ ; -10^\circ < b < +5^\circ)$ and consequently we
do not presently reddening correct our fields with $b<-10^\circ$.  We hope in the near future to develop a similar approach to \citet{simion17} by employing our BDBS $iz$Y band to reddening correct our complete dataset.   The transformations from the Vista system to 2MASS and individual total extinction corrections for each of our filter bandpasses are given in Section 3.4 of \citet{bdbs2}, where the reddening is as given in \citet{green18}; these values were derived using disk and bulge stars. We adopt the $u$ band total extinction 
for $R_v=3.1$ as given in Table 6 of \citet{sch11}.  At this time we are not undertaking a study of the reddening law and its variation across the field.

\subsection{Survey footprint characteristics}

\begin{figure*}
\centering
\includegraphics[scale=0.3]{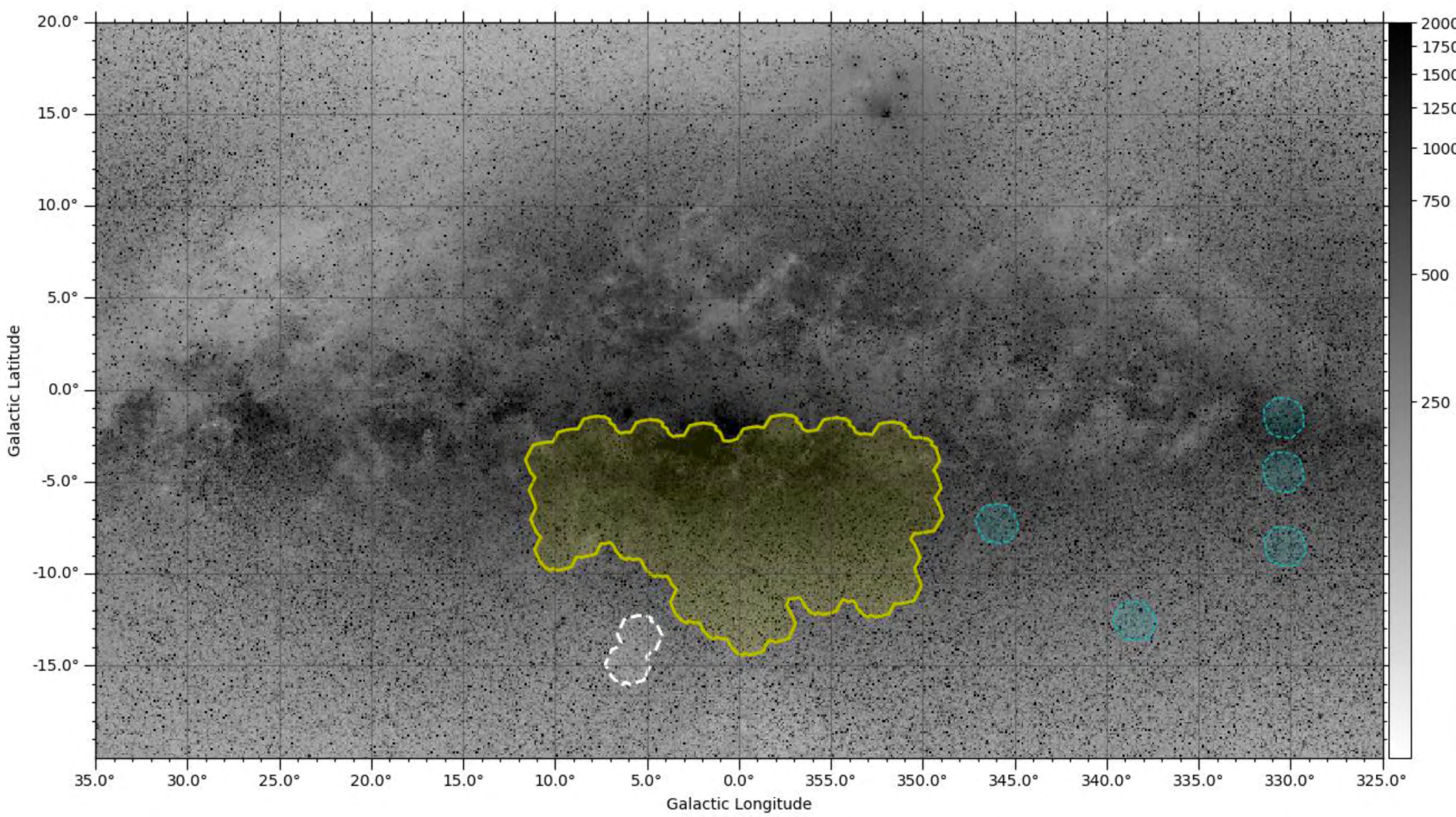}
\caption{The BDBS footprint including the extension placed on the core of the Sgr dwarf spheroidal galaxy, superimposed on a greyscale map of the Milky Way from \citet{mel09pano}.   The field enclosed in yellow is fully reduced and calibrated.  Other fields are illustrated, but they are not yet reduced.  The two joined white circles are chosen to sample the core of the Sagittarius dwarf spheroidal galaxy.  The fields indicated in blue were selected to sample fields in the disk, to be used for eventual statistical subtraction of the foreground disk population.}
\label{bigfoot}
\end{figure*}

Figure \ref{bigfoot} shows our final survey footprint superimposed on an optical image of the Galactic bulge \citep{mel09pano}.  As ours is an optical survey, we avoid regions of high extinction in the plane and at $b>0^\circ$.  However, if maps of the extinction can be produced, the Northern bulge is potentially accessible using the techniques described herein and in our companion paper \citep{bdbs2}.  The two fields that are detached from the main footprint are chosen to fall in the main body of the Sagittarius dwarf spheroidal galaxy.  These were originally observed under non-photometric conditions and we have only recently obtained calibration images in photometric  conditions; photometry of these fields will be reported in a future paper.  

Planning to eventually employ statistical subtraction of the foreground disk population, we also added five additional fields intended to sample the disk population at $l=-30^\circ$ spanning $-2^\circ <b< -8^\circ$; these are also shown in \ref{bigfoot}.  We use the approximate $(l,b)$ for these fields and approach successfully advanced in \citet{ort95} and \citet{zoc03}.  The control fields are designed to sample the foreground disk in regions where the bulge makes little contribution, and to provide a sample for statistical subtraction when we address the age of the bulge. 

As \citet{shapley1918} first noted, most of the globular clusters in the Milky Way are found toward the Galactic Center.  The BDBS field includes 26 globular clusters; these will be used to provide red giant branch templates to calibrate metallicities across the bulge, and their properties will be studied as well, using our dataset.   It is also noteworthy that maps of the X-structure, especially those in \citet{saito11} and \citet{wegg13}, show that the bulge x-structure is most prominent for $\lvert l \rvert <1^\circ$.  Our footprint is designed to overlap the full extent of the X-shaped bulge.  Table 2 below gives a list and salient properties of globular clusters that fall within the BDBS footprint. 

\begin{table*}
	\centering
	\label{tab:table2}
    \tabcolsep=0.2cm
    \footnotesize
    \caption{Globular Clusters in the BDBS Footprint.$^a$}
   \begin{tabular}{lllllcccl}
    
\hline
Cluster & RA (2000)  & Dec (2000) & $l$ & $b$ & $M_v$ & [Fe/H]  \\
   
\hline

FSR1758 $^b$     
& 17  31 12 &   -39 48 30    &    262.80  &  -39.81 & -8.6$^c$   & -1.50 \\

NGC 6380	&  17 34 28  &   -39 04 09  &   350.18  &  -3.42  &  -7.5 &  -0.75 \\

Ton2  &	17 36 11 &  -38 33 12  &   350.80 &  -3.42  &  -6.17 &  -0.70 \\

Djorg1  &	17 47 28 &    -33 03 56  &   356.69  &  -2.47  &  -6.98  &  -1.51 \\

NGC 6441   &  17 50 13 &  -37 03 05 &   353.53  &  -5.01 &  -9.63 &  -0.46 \\

Terzan6 &	17 50 31 & -31 19 30  &     267.63 & -31.32  & -7.59 &  -0.56 \\

NGC 6453	& 17 50 52 &   -34 35 57  &  355.72  &  -3.87 &  -7.22 & -1.50  \\

Terzan9	&  18 01 39  &  -26 50 17 & 270.41  &  -26.84  &  -3.71 &  -1.05 \\

Djorg2 	& 18 01 49 &   -27 49 33   &    2.77  &  -2.50  &  -7.00  &  -0.65 \\

NGC 6522	& 18 03 34  &  -30 02 02  &    1.02  &  -3.93  & -7.65 &  -1.34 \\

Terzan10 & 	18 03 39 &  -26 05 08 &	270.91 & -26.08  &   -6.35 &  -1.00\\

NGC 6528	& 18 04 50 &  -30 03 22 &    1.14 &  -4.17 &  -6.57 &  -0.11 \\

NGC 6540	& 18 06 09 &  -27 45 55  &      3.29 &   -3.31 &  -6.35 &  -1.35 \\

NGC 6544 &	18 07 21  &  -24 59 50 &    5.84 &   -2.20 &   -6.94 &  -1.40 \\

NGC 6553 &	18 09 18  &  -25 54 31  &  5.26  &  -3.03  &  -7.77 &  -0.18 \\

NGC 6558 & 	18 10 18  &  -31 45 50  &    0.20  &  -6.02 &  -6.44 &  -1.35 \\

Terzan 12	 & 18 12 16   & -22 44 31   &     8.37  &  -2.10 &  -4.14 &   -0.50  \\

NGC 6569	& 18 13 39  &  -31 49 37 &    0.48  &  -6.68 &   -8.28 &  -0.76 \\

AL3 = BH261  & 18 14 07 &  -28 38 06 &   3.36 &  -5.27  & ... & ... \\      

NGC 6624	&  18 23 41 &  -30 21 40  &    2.79  &  -7.91 &  -7.49  &  -0.44 \\

NGC 6626	& 18 24 33 &  -24 52 11   &   7.80   &  -5.58  &  -8.16  & -1.32 \\

NGC 6637	& 18 31 23 &  -32 20 53  &    1.72 &  -10.27 &  -7.64 &  -0.64 \\

NGC 6642       & 18 31 54 &  -23 28 31 &  9.81  &  -6.44 & -6.66 & -1.26 \\

NGC 6638	& 18 30 56 &  -25 29 51   &   7.90 &   -7.15 &  -7.12 &  -0.95 \\

NGC 6652       & 18 35 46 &  -32 59 27 & 1.53  & -11.38 & -6.66 & -0.81 \\

NGC 6656	& 18 36 24 &  -23 54 17  &    9.89 &   -7.55   &  -8.5 &  -1.70  \\
\hline

\multicolumn{7}{l}{$^a$  Parameters from the Harris catalog of globular cluster parameters \citet{harris10} } \\
\multicolumn{7}{l}{$^b$ parameters from \citet{barba19} }\\
\multicolumn{7}{l}{$^c$ magnitude reported is $M_I$}\\
\end{tabular}
\end{table*}

\begin{figure*}
\centering
\includegraphics[scale=0.7]{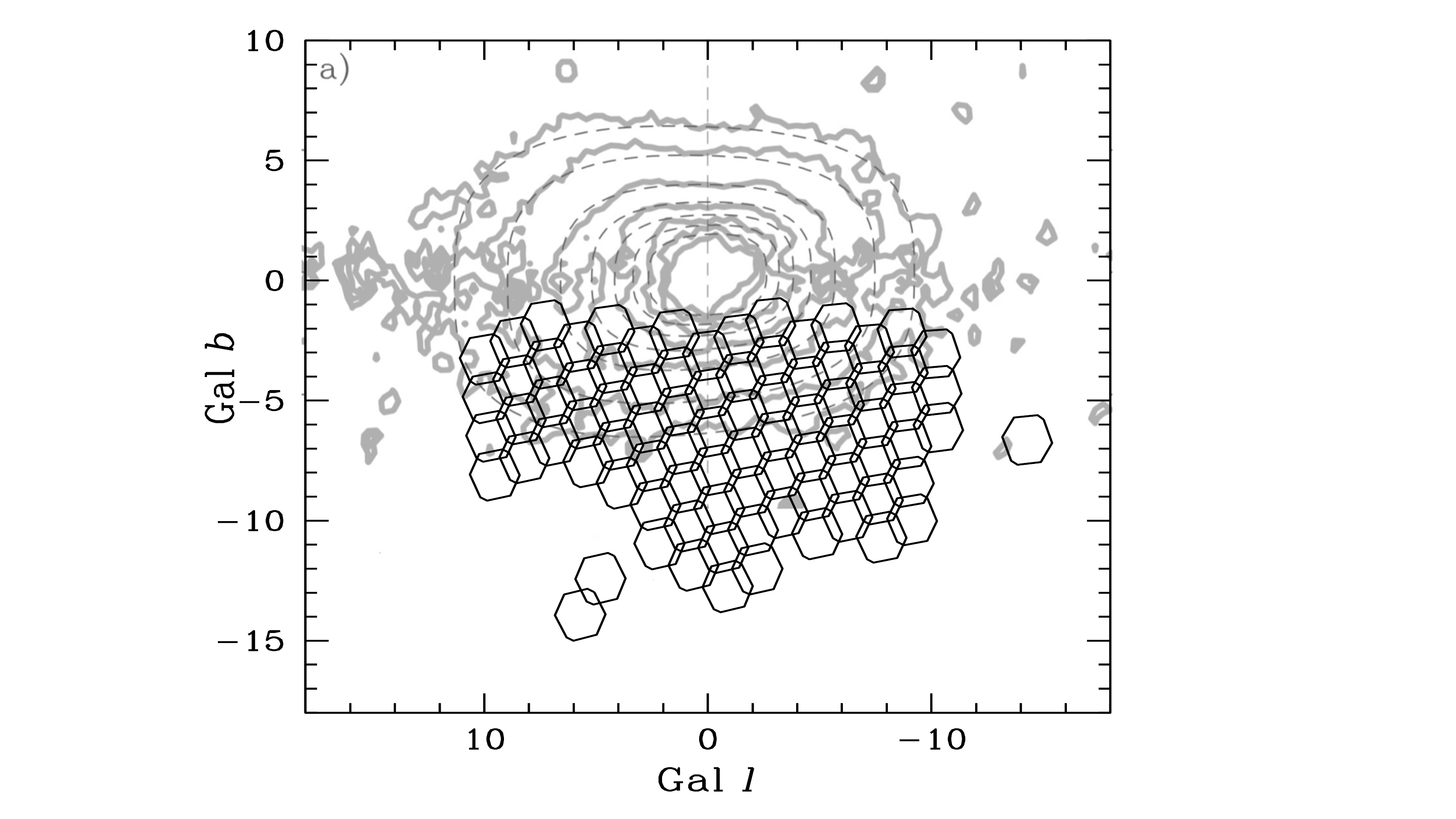}
\caption{The BDBS fields are overlaid on the 2.2$\mu \rm m$ map of the Milky
way from \citet{launhardt02}.  The footprint is designed to sample a large fraction of the mass of the bulge. The contours are the weighted average of maps at 2.2, 3.5, and 4.9 $\mu \rm m$,scaled to the 2.2$\mu \rm m$  surface brightness}
\label{bdbs-laun02}
\end{figure*}

We now consider how our survey overlays the infrared light of the bulge.
The infrared map ( Figure \ref{bdbs-laun02}) of \citealt{launhardt02} shows that most of the  2$\mu \rm m$ luminosity of the Galactic bulge is encompassed in a $\pm 10^\circ$
boundary in Galactic longitude and latitude.  BDBS samples nearly fully this field of regard and adds two fields placed on the core of the Sgr dwarf spheroidal galaxy.   The initial survey reaches close to the plane at $b=-2^\circ$ but stops there due to the extremely high optical extinction. We envision spanning the footprint of the VVV survey and to extend BDBS to the bulge North of the plane.  However, as BDBS is an optical survey, correction for the variable and high extinction North of the plane will be challenging.

%template for column width figure
%\begin{figure}
%\centering
%\includegraphics[width=\columnwidth]{bdbs-footprint.png}
%\caption{footprint}
%\label{footprint}
%\end{figure}

\begin{figure*}
\centering
\includegraphics[scale=0.7]{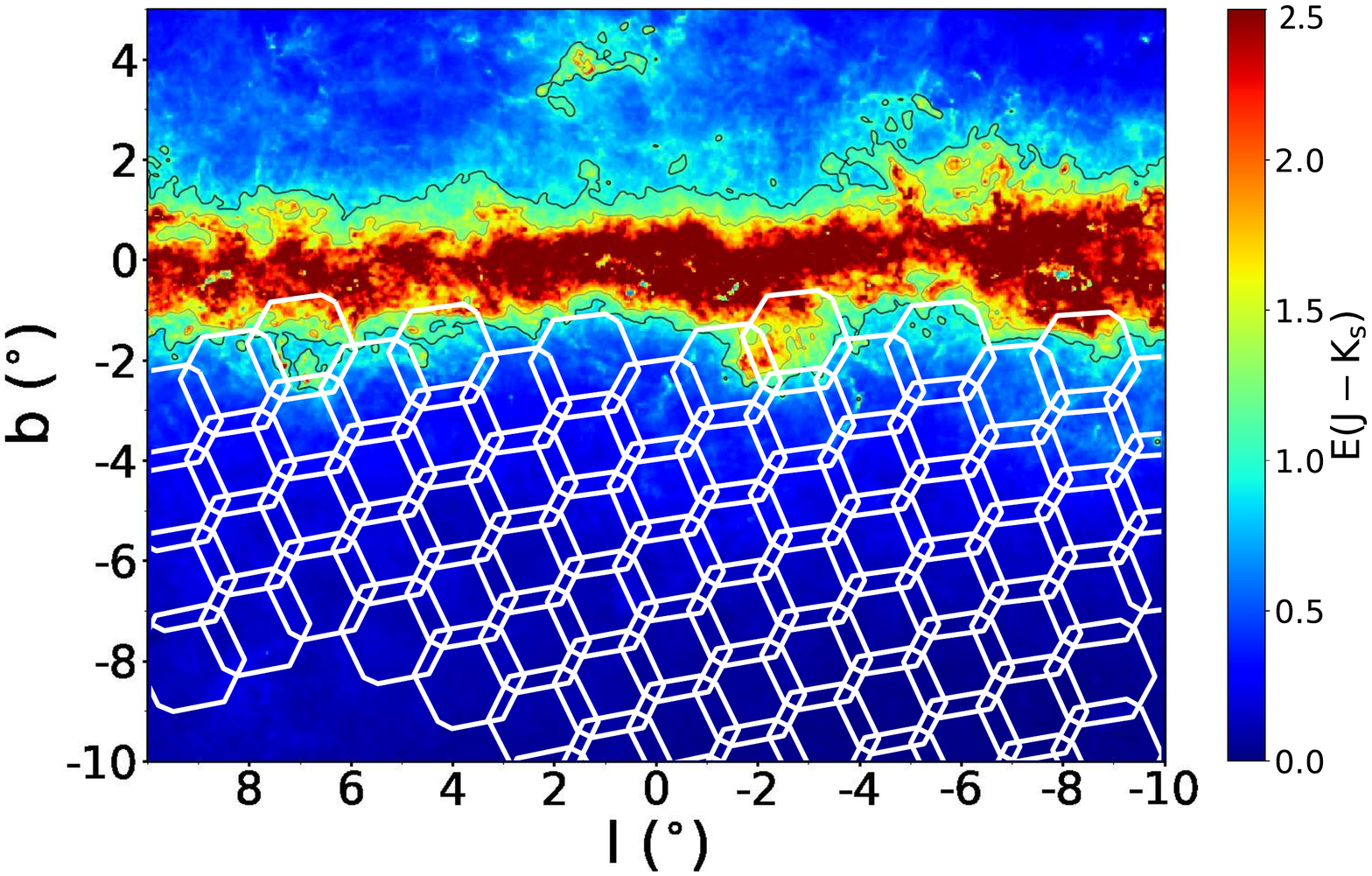}
\caption{The BDBS fields are superimposed on the reddening map constructed from $1\times1$ arcmin cells \citep{simion17}.  At this point, the
reddening map is based on data from the VVV survey; we ultimately plan to deredden over the full BDBS footprint.}
\label{bdbs-redden}
\end{figure*}

Figure 3 shows the BDBS footprint superimposed on a map of RA, DEC and $l,b$, and also reddening, based on the $1\times1^\prime$ reddening map from \citet{simion17}.  Our survey limits exceed those of VVV and consequently not all of  BDBS is included in the current reddening map.  Periodic updates of the BDBS database will include dereddening for the full field, and improved reddening.  As we mentioned earlier, BDBS avoids regions closer to the plane than $1^\circ$; in such regions of high extinction, only infrared photometry will suffice for investigating the bulge.  However, the BDBS footprint takes us within 140 pc of the plane and therefore is able to well sample the bulge for exploring properties such as age and metallicity.

\subsection{Astrometry with BDBS and Gaia}

\begin{figure*}
\centering
\includegraphics[scale=1]{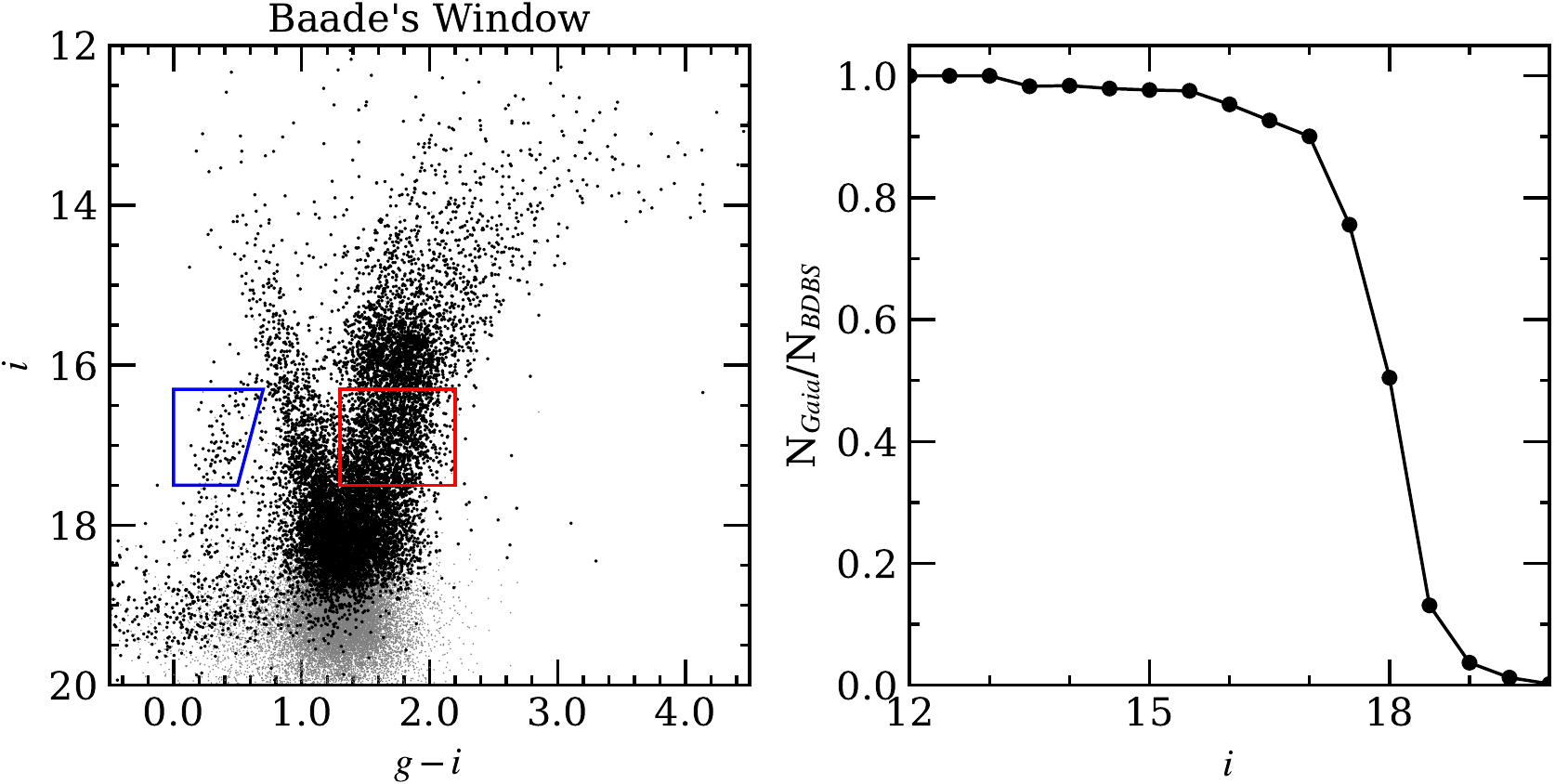}
\caption{(Left): A colour-magnitude diagram of Baade's Window $(l,b)=(0.9^\circ, -3.9^\circ)$ in the Galactic bulge with selection boxes indicating blue HB and RGB stars. (Right): The fraction of stars with Gaia DR2 matching, as a function of $i$ magnitude.  The completeness limit for Gaia DR2 matching is substantially fainter than the red clump, in this field.}
\label{bdbs-Gaia}
\end{figure*}

A critical aspect of our survey will be the use of Gaia DR2 astrometry to obtain kinematics as a function of age and metallicity.  Figure \ref{bdbs-Gaia} shows that our Gaia DR2 completeness is excellent to $i=18$ in Baade's Window at $(l,b)^\circ = (0.9,-4) $.  In future data releases, it may be possible for matched Gaia DR2 detections to reach the main sequence turnoff over much of the BDBS footprint, making possible the veto of foreground stars from analyses of the bulge age.

BDBS offers the possibility to perform scientifically useful astrometry at fainter magnitudes (and thus to more spatially crowded regions) than Gaia DR2 can reach towards the bulge. A full accounting of the astrometric precision achieved with DECam is a substantial undertaking (even in uncrowded regions, e.g. \citet{bernstein17}, which demonstrates absolute accuracies of $\sim$7mas at magnitudes for which Gaia DR2 furnishes a sufficiently dense reference catalog). Detailed analysis of our achieved astrometric precision with BDBS is still in progress. For rough indications, however, we use the S/N actually achieved \citep{bdbs2} and thus accounting for spatial confusion: the ultimate astrometric precision delivered by BDBS will be worse than these early indications. 

We adopt photometric S/N$\sim$ 20 as a sensible limit to check completeness to useful astrometry: at this photometric precision, simple scaling with the PSF FWHM suggests per-epoch astrometric repeatability on the order of 30-50 mas per epoch, roughly in line with an extrapolation of the Gaia DR2 delivered astrometric uncertainty curve down to $r \sim 19-20$.  The relatively early epoch of observation of BDBS extends the time baselines of future planned surveys (such as the 10-year LSST survey).  Thus, proper motion precisions of a few milliarcsec/yr become feasible: on a population basis, this opens up the possibility of kinematically tagging bulge- and disk- populations.  

The rewards in survey grasp are substantial. In the 0.25 degree radius field within Baade's Window, for example, the extra depth leads to 96,000 more detected objects in BDBS than for Gaia DR2 (more than doubling the sample size). This increase in detected population is typical for the fields surveyed to date.  

Note that we are not claiming that BDBS outperforms Gaia DR2 astrometrically (space-based astrometry generally being far superior to ground-based at the same S/N). However, we do assert that DECam is capable of achieving scientifically useful astrometry down to about 1-2 magnitudes fainter than Gaia DR2 in most fields in the inner Bulge. For Baade's Window, for example, BDBS reaches limiting magnitude $r \approx 20.5 (21.8)$~at 0.05 mag precision in r (g) filters (assuming S/N selection on r and g only for a direct comparison to Gaia DR2; when the five filters g,r,i,z,y are considered in the selection, the limits are about 0.5 mag brighter. This limit is starting to reach the main sequence turn-off in this field; when more specialized analyses are brought to bear, we expect that the magnitude limits can be pushed still fainter - possibly to the middle of the MSTO in some fields. 

\begin{figure*}
\centering
\includegraphics[scale=1]{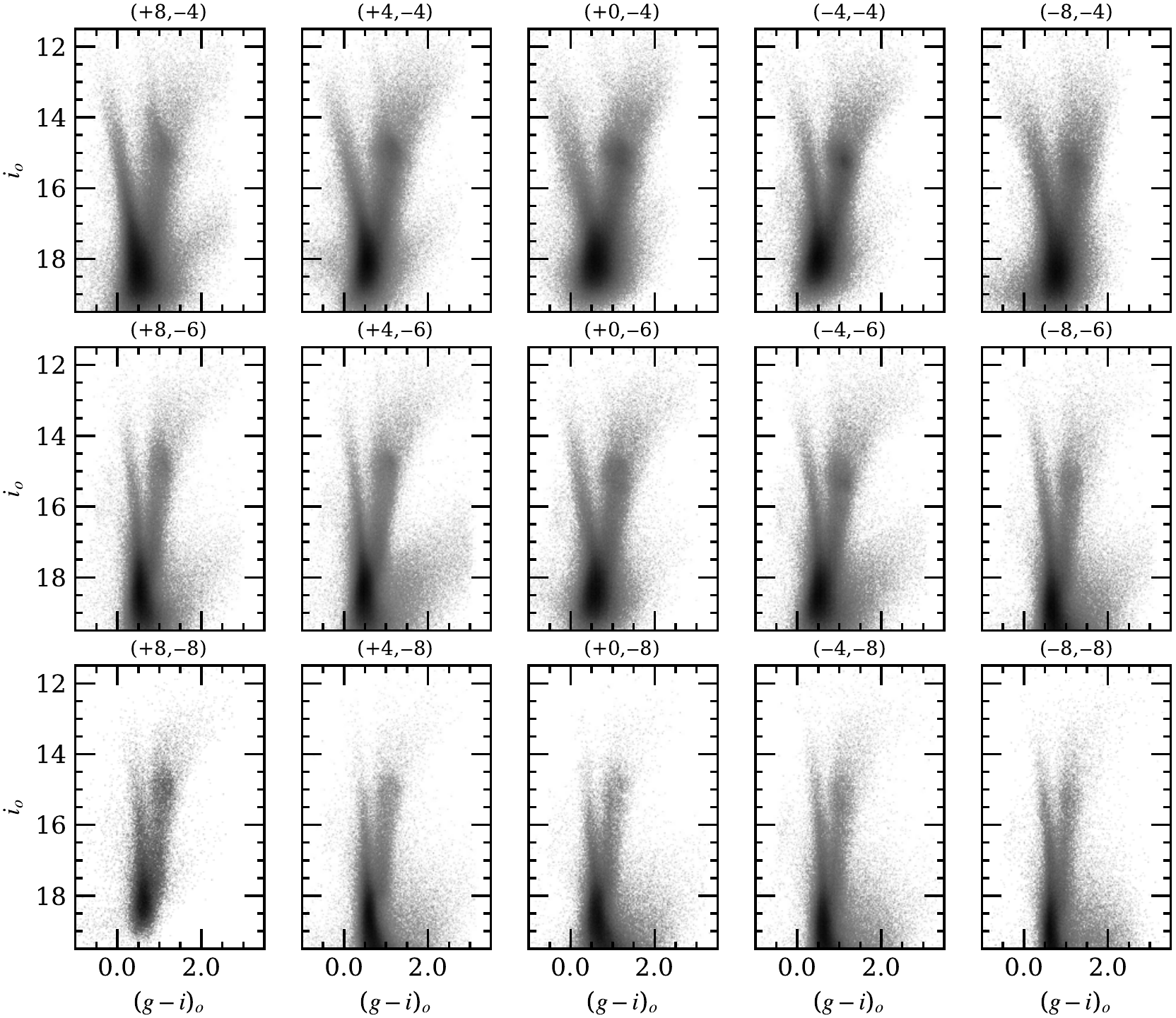}
\caption{Fig. 5-Panel of dereddened CMDs spanning the bulge, sorted by $(l,b)$.  Notice that the form of
the red clump is peculiar in the region toward $(l,b)=+8^{\circ}, -4^{\circ}$ that likely reflects the presence of the
"long bar" population.  The doubling of the red clump is clearly evident at $(l,b)=0^{\circ}, -8^{\circ}$}
\label{bdbs-cmdpanel}
\end{figure*}

\section{Early Results Concerning the Stellar Population}

\subsection{The bulge population}

Figure \ref{bdbs-cmdpanel} presents an array of colour-magnitude diagrams spanning the Galactic bulge, in $(g-i)_0$.  All fields include a blue foreground disk star sequence, a red giant branch, red clump, and bulge main sequence turnoff point.  We have dereddened our data using the 1 arcmin scale reddening map derived from the VVV data by \citet{simion17}.  

Most of our bulge fields also include an extended blue horizontal (or "vertical") branch, although in some fields, a globular cluster can be the majority contributor to the BHB.  This is true especially in Baade's Window, where the metal poor globular cluster NGC 6522 contributes a large fraction of the BHB in that field (e.g. \citet{saha19}).

We can immediately note some significant features in the colour-magnitude diagrams.  First, the foreground disk is ubiquitous in all fields. This is a main sequence structure that presents as a blue linear feature in the CMD.  It is also evident that the red giant branch is both more broad and redder in the lower latitude fields.  While this might be attributed to increased differential reddening, the difference between fields at $-4^\circ$ and $-8^\circ$ is real as both are dereddened.    The metallicity distribution is discussed in \citet{bdbs2}, but one can visually notice that the RGB becomes both more narrow and vertical in the fields at $-8^\circ$, reflecting a real decline in metallicity compared to the lower latitude fields.   Also, at $b=-8^\circ$ and $l<0$ one can see clearly the doubling of the red clump, which is attributed to an X-structure in the bulge.  We will investigate this feature and its structure as a function of metallicity, in future work. 

The peculiar appearance of the red clump at $(l,b)=+8,-4$ is noteworthy. The red clump appears to have a bright extension that we speculate is likely caused by red clump stars in the disk, lying closer to the Sun that the bulk of the bulge.  It is also possibly due to the "long flat" bar (\citet{lopez07},\citet{cabrera07}, \citet{cabrera08}); future modeling is needed to choose between these alternatives.

\subsection{Is there a Young, Massive Population in the Bulge?}

As mentioned previously, the large contiguous BDBS field is ideal for exploiting the Gaia DR2 database.  The extinction is low enough, and the bulge population bright enough, to have significant overlap with Gaia DR2, which we use to resolve some interesting problems in the interpretation of the bulge population.

\begin{figure*}
\centering
\includegraphics[scale=0.8]{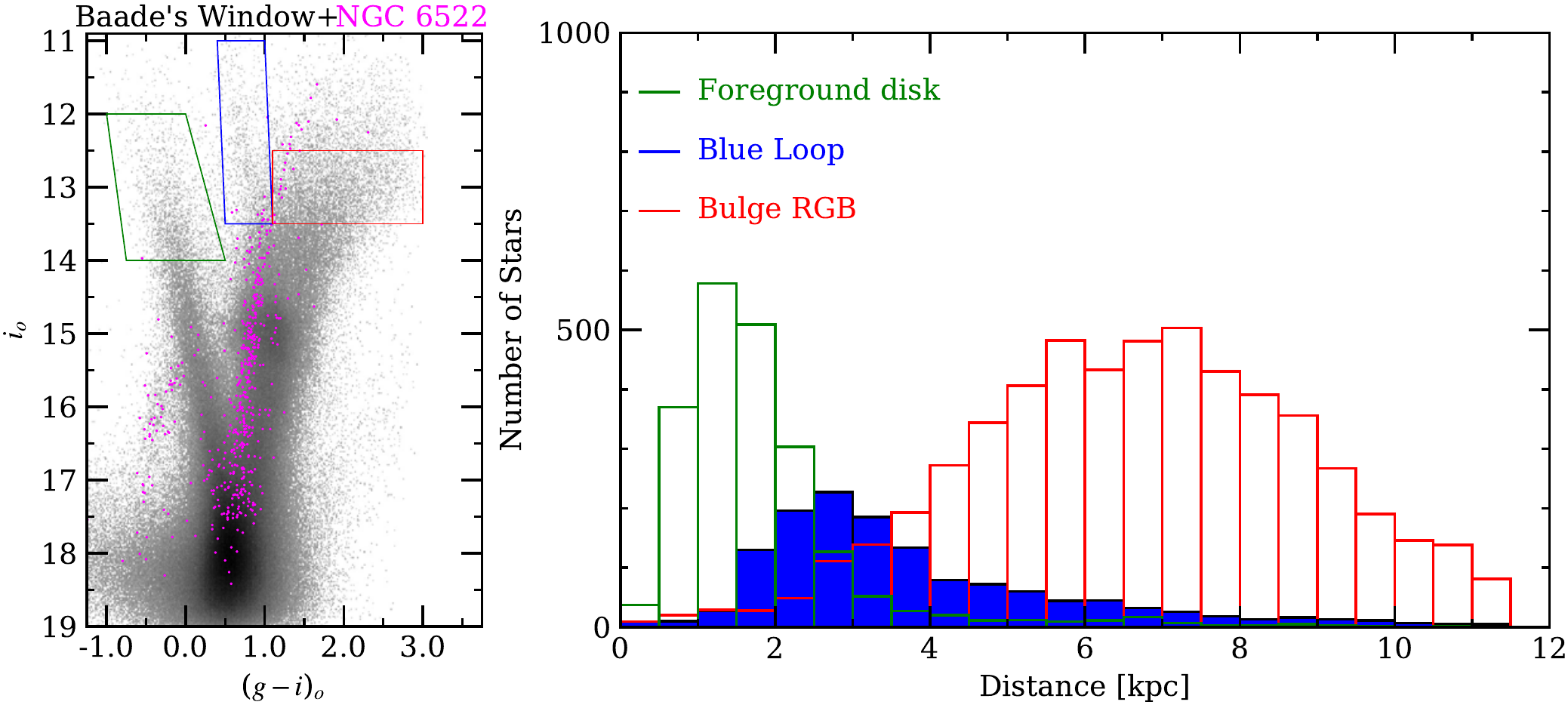}
\caption{The colour-magnitude diagram of Baade's Window is analyzed with distances of foreground disk,  candidate massive blue loop stars \citep{saha19}, and bulge RGB stars from
Gaia DR2 \citep{bailerjones18}.  Members of NGC 6522 are illustrated in pink.  Notice that all stars are placed at their expected distances, with the foreground disk stars lying mostly with 2 kpc.  We find the young blue loop stars proposed by \citet{saha19}  to lie mostly in the distance range of 2-4 kpc; we propose that these are late-type stars (red clump/giants) that belong the thin disk or bar, and are closer to the Sun than the bulge.}
\label{bwdistances}
\end{figure*}

\citet{saha19} in their Figure 1 propose that stars in the blueward/bright extension of the red clump are massive helium burning stars in the bulge, in the advanced "blue loop" phase of stellar evolution.  This feature in the colour-magnitude diagram is clearly visible in many fields shown in Fig \ref{bdbs-cmdpanel}.  It is important to understand the cause of this feature, as it is present over most of our field, especially at lower Galactic latitudes. We test Saha's hypothesis in Figure \ref{bwdistances} by matching this feature to the \citet{bailerjones18} catalog, and we find that nearly all of these stars have derived distances $\sim 1.5$ kpc from the Sun; this would make their colours and magnitudes consistent with red clump stars lying closer to the Sun than the bulge, as opposed to a population of young, massive stars.   We confirm this finding for other methods of distance determination using the Gaia DR2 catalog, including using only the parallax data.   If one imagines displacing the bulge red giant population closer to the sun, as would be the case if the clump were in the disk, the intrinsic magnitudes and colours would trend along a track that is brighter and bluer. Further, stars lying in {\it front} of the reddening sheet would be (wrongly) corrected for reddening.  We propose this interpretation for the blue/bright red clump feature as an alternative to the massive star population, and it appears to account for this peculiar feature in the CMD.  

\begin{figure*}
\centering
\includegraphics[scale=0.75]{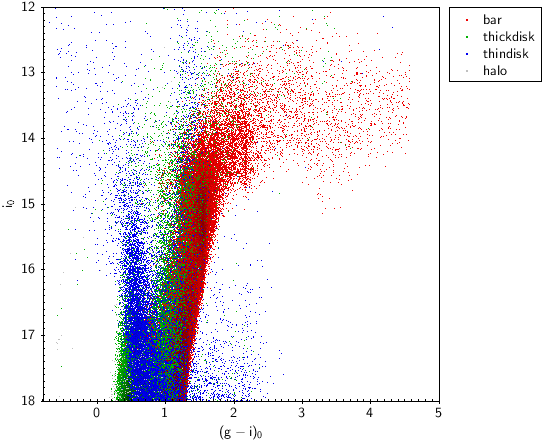}
\caption{A model CMD of the population toward Baade's Window $(l,b)=(-0.2^\circ, +3.9^\circ)$ covering 0.1 sq. deg. (see text and the Besancon Galaxy Model; \citet{lagarde17}). This model breaks down the the CMD by stellar populations as indicated in the legend.  Notice that the bright blue population near the red giant branch is dominated by stars in the thin disk, not the bulge.}
\label{robinfigcmd}
\end{figure*}

\begin{figure*}
\centering
\includegraphics[scale=0.60]{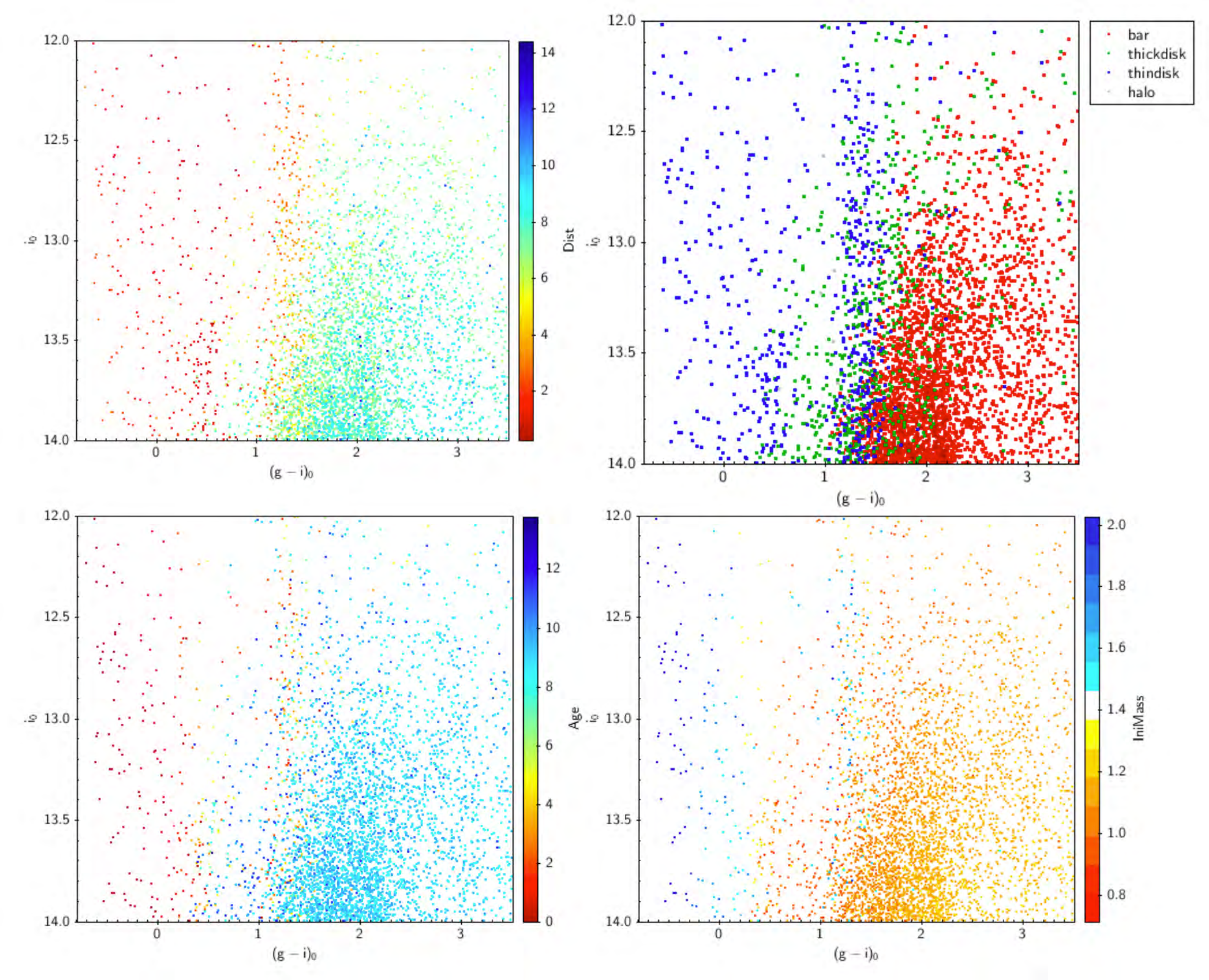}
\caption{A model CMD of the population toward Baade's Window $(l,b)=(-0.2^\circ, +3.9^\circ)$ covering 0.1 sq. deg. (see text and the Besancon Galaxy Model; \citet{lagarde17}). Clockwise from the top are panels showing the distance in kpc, the  population (indicated in the legend), the stellar mass, and the age in Gyr.  The population proposed by \citet{saha19} as the massive blue loop stars would be the blue (disk stars) in the population panel, near $i=12.5$ and $g-i=1$.  This model would assign the blue plume"to the foreground disk population, which might plausibly explain the unusual bright extension seen in the red clump near $(l,b) = (+8^\circ, =4^\circ)$ that is seen in Fig \ref{bdbs-cmdpanel}.}
\label{robinfig}
\end{figure*}

The feature is especially prominent toward the plane, as would be expected if the stars arise in the old disk lying along the line of sight toward the bulge.   A different approach confirms our interpretation using the Besancon model \citep{robin12a,robin12b} and concurs that the proposed blue loop stars are actually older than 1 Gyr, but reside in the foreground disk.   A new model has been used to simulate a  0.1 sq. deg. 
field toward Baade's Window $(l,b)=(-0.2^\circ, +3.9^\circ)$ and the resulting CMD is show in \ref{robinfigcmd} and \ref{robinfig}.   The CMD was generated using the model version mev1910, that uses new evolutionary tracks as described in the Besancon Galaxy Model \citep{lagarde17,lagarde19}; the bar model is described in \citet{robin12b}, with modifications of the mass density from Awiphan et al. (in prep).    
           
Considering that a population of such young massive stars would be expected
to have luminous evolved counterparts, we prefer our conclusion that these are red clump stars in the foreground, as opposed to young, massive stars at the bulge distance of 6-8kpc; they are indeed identified to lie at distances where we have high confidence in the Gaia DR2 parallaxes.

\subsection{A Peculiar Red Clump:  Long Bar signature, or disk feature?}

\begin{figure*}
\centering
\includegraphics[scale=1.0]{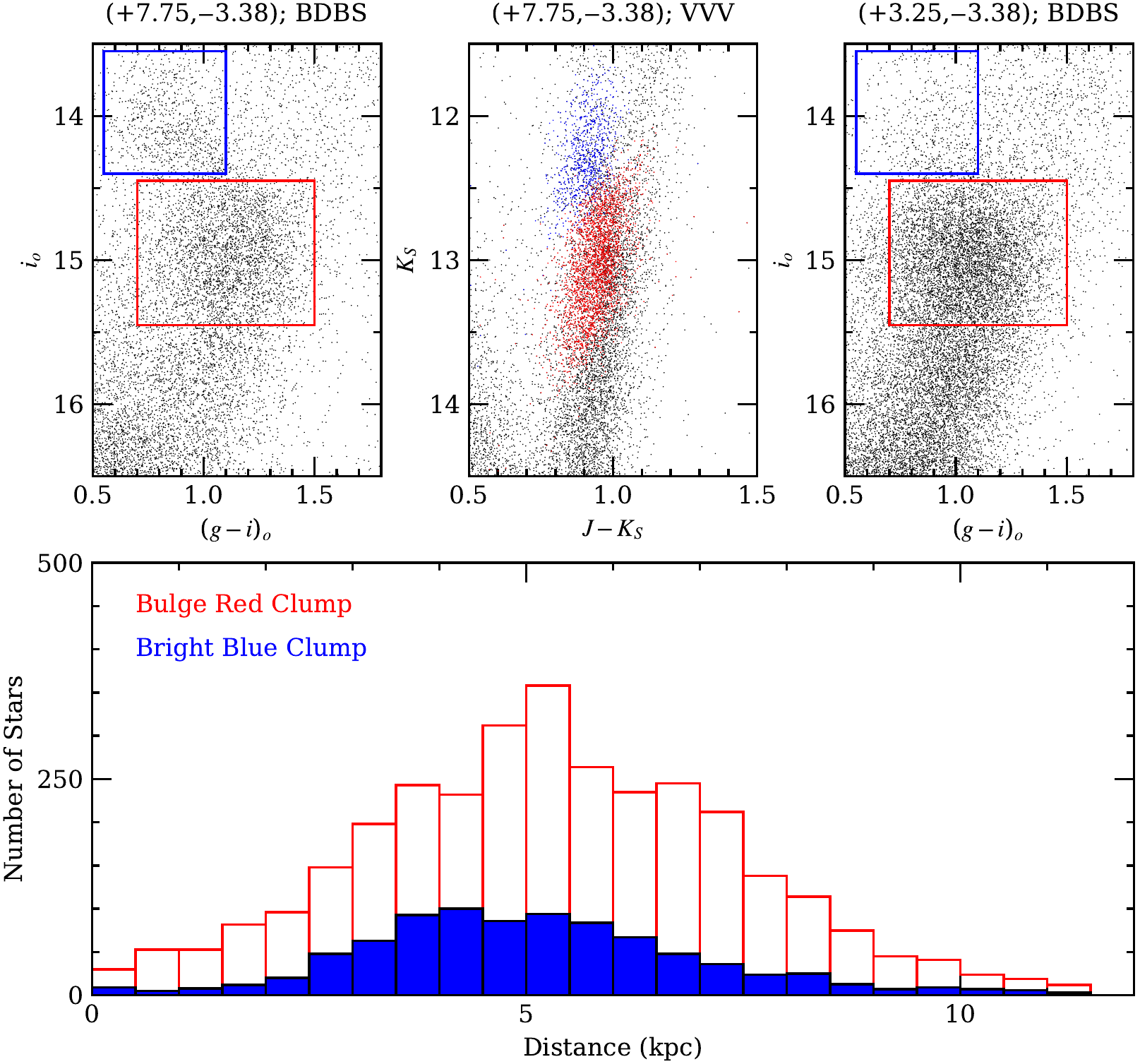}
\caption{Stars in what appears to be a foreground red clump population are examined.  The clump region is divided into a fainter bulge population and brighter (blue) candidate foreground population.  Gaia DR2 distances \citep{bailerjones18} suggest the bright blue feature consists of stars that are indeed, closer to the Sun than the bulk of the bulge red clump stars.  This feature may arise in the foreground thin disk (Fig \ref{robinfig}), or the "long bar".}
\label{bar_clump}
\end{figure*}

It has been proposed (e.g. \citet{lopez07}) that toward the plane, the bar is both more flattened and extended; this proposed feature is designated as the "long bar".
Figure \ref{bar_clump} shows Gaia DR2 distances for the field at $(l,b)=+7.75,-3.4)$ in VVV and BDBS.  The feature first noted by \citet{lopez07} is clearly seen in our field and diminishes as one moves toward the minor axis.  We can examine the Gaia DR2 derived distances for stars in the red clump extension, and note that they are somewhat closer than the fainter red clump that we associate with the main bulge population.   Figure \ref{robinfig} presents the case that bright bar extension is actually due to stars in the foreground disk.  

Fortunately, the stars in question are bright enough that improved parallaxes and proper motions can be expected during the GAIA mission.  High resolution spectroscopy might also shed light on the nature of this population by enabling a self consistent measurement of the spectroscopic parallax and metallicity.  At present, combining the colour-magnitude diagram and GAIA-derived distances suggests only that the bright blue clump is just marginally closer than the red clump.

\subsection{ Mapping the blue horizontal branch}

The wide field of regard also enables mapping of the spatial distribution of stellar populations in the bulge colour-magnitude diagram.  Our long term plan will be to combine Gaia DR2 data with the optical photometry spanning the whole field; this will become feasible with anticipated improvements that should come with Gaia DR3.

We turn to one of the most interesting features of the CMD, the blue horizontal branch.  BHB stars are a signature of old stellar populations; they once were thought to arise only from metal poor populations but are now known in metal rich populations as well, and may be the stars responsible for the as yet unsolved UV rising flux of elliptical galaxies.  BHB stars are of special interest, because they are also potential tracers of dissolved globular clusters and dwarf spheroidal galaxies.  A disrupted system that is rich in BHB stars will leave a clear signature in such a map--either clump or stream-like features.  Finally, clusters of BHB stars can in principle point the way toward previously uncatalogued globular clusters.

We create a uniform spatial grid across the entire BDBS footprint for which dereddening values from VVV \citep{simion17} are available.   The boundary is therefore the VVV footprint, not the BDBS footprint.  Within each grid point a dereddened $i$ vs $g-i$ CMD is created; we generate a selection box around the region that maximizes the number of blue HB stars and minimizes contamination from the foreground disk (e.g., see Fig \ref{bdbs-Gaia}), and we calculate the number of likely blue HB stars in the selection box.  The resulting colour map reflects the number of blue HB stars within each grid point.  The open circles signify the positions of known globular clusters.  
\begin{figure*}
\centering
\includegraphics[scale=1.0]{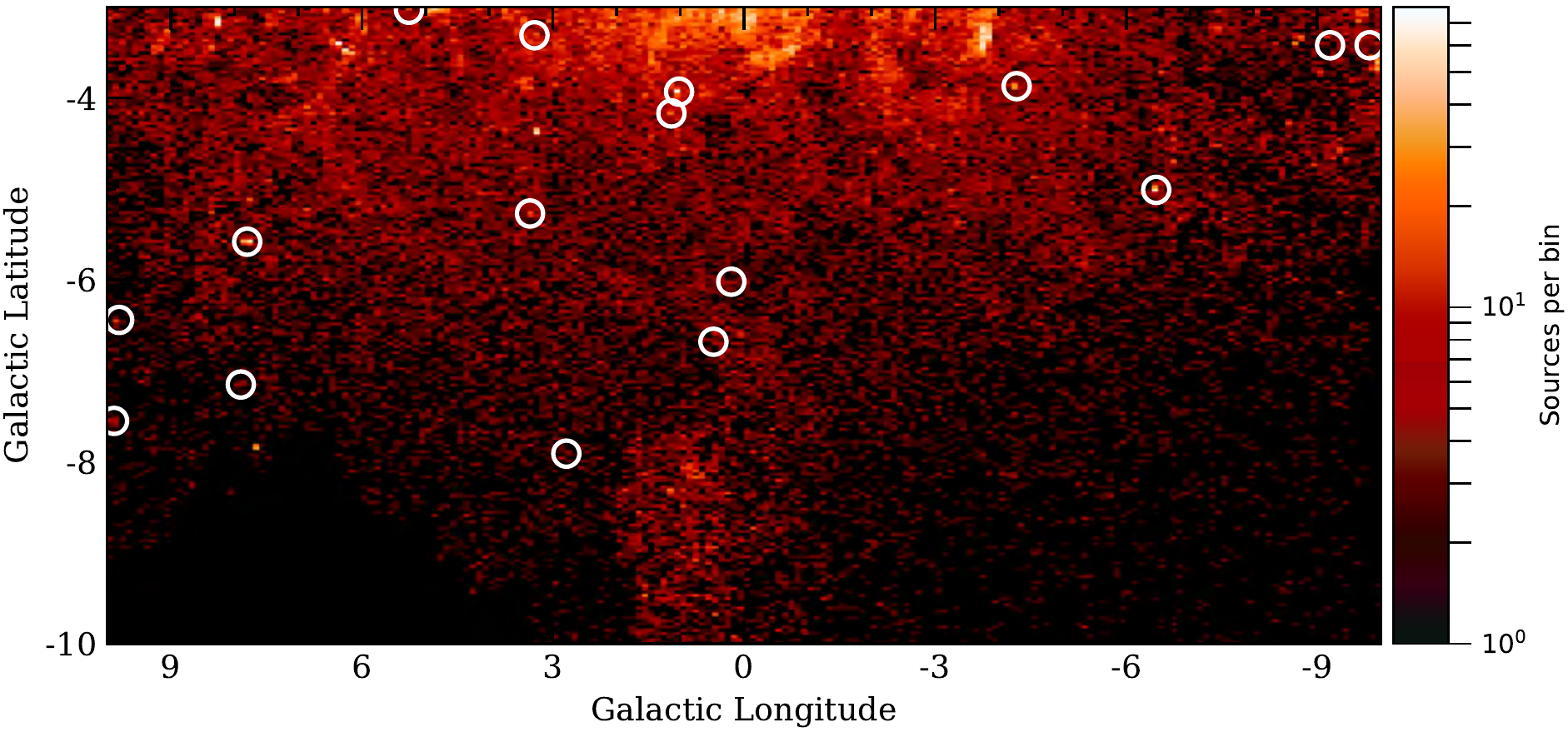}
\caption{A preliminary map of the density of blue HB stars, with superimposed black circles indicating the locations
of known globular clusters.  All concentrations of BHB stars correspond either to known globular clusters or approximately with the Sgr dwarf spheroidal galaxy.  The irregular 
features near the plane do not appear to correspond to actual concentrations.  A modest increase in density of BHB stars is noted near $(l,b)=+2,-9$ that corresponds with an extension of the Sgr dwarf spheroidal galaxy.  White circles indicate the positions of known globular clusters. }
\label{bhbzoom}
\end{figure*}

\begin{figure*}
\centering
\includegraphics[scale=1.0]{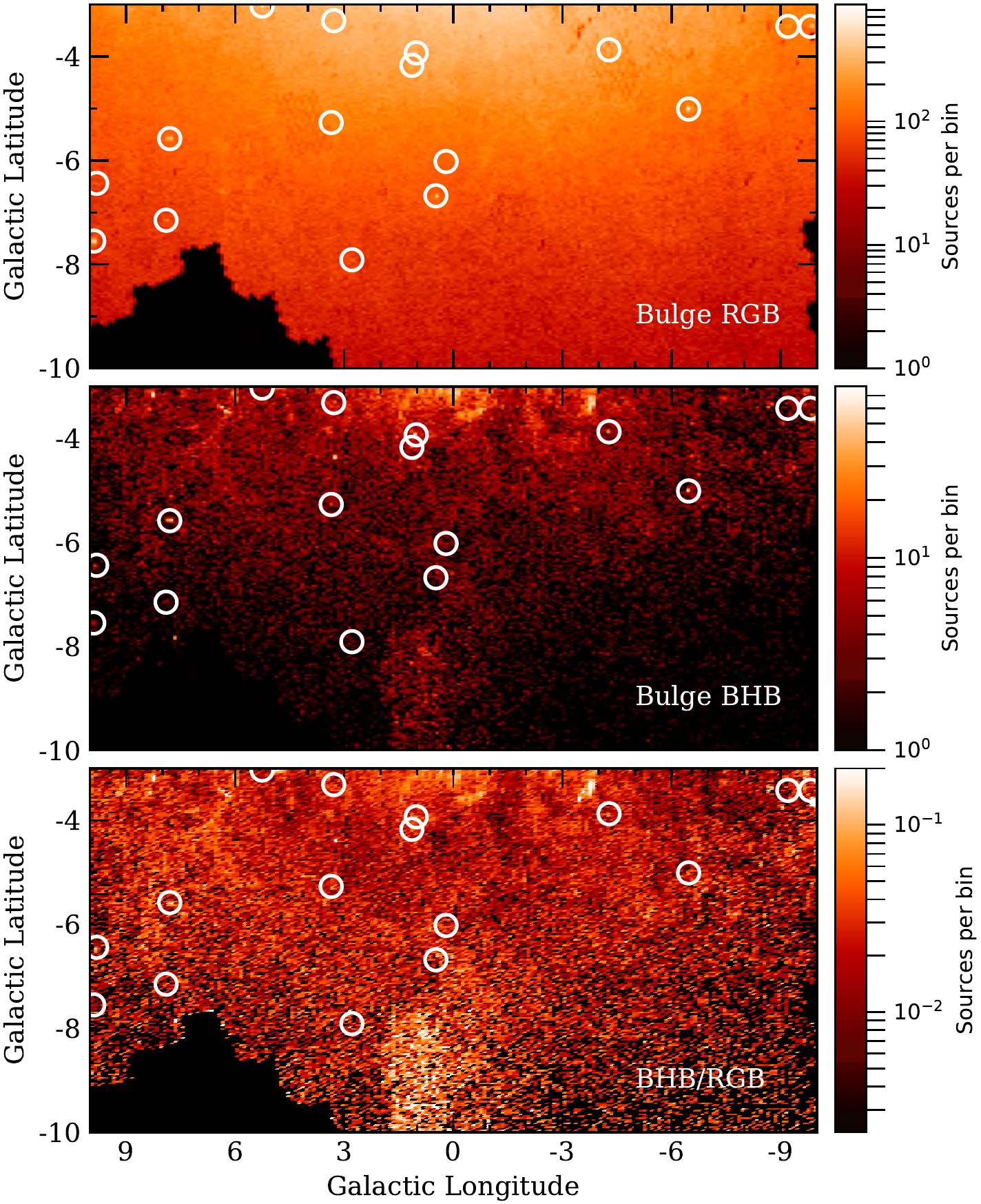}
\caption{Mapping the blue HB of the Milky Way.  Upper panel is the density of RGB stars; center panel is the density of BHB stars as in Fig \ref{bhbzoom}; the bottom panel is the ration of BHB/RGB stars.  }
\label{bhbplot}
\end{figure*}

Figure \ref{bhbzoom} shows our first efforts at mapping the density of the BHB.  As the BHB extends to the limit of our detection as defined by crowding, we can consider these density maps to be only approximate, and most useful as a guide to features that may be interesting for future investigation.  However, in this figure one can clearly see that all of the significant peaks of density in the BHB population are associated with known globular clusters, and no stream-like features are evident.   We suspect that some of the variation near the plane is due to variable reddening and completeness, rather than an actual reflection of true population variation. There is a interesting apparent enhancement in BHB/RGB at $(l,b)=+2^\circ, -9^\circ$ that is roughly in the direction of the Sgr dwarf spheroidal galaxy
(refer to the Sgr dwarf fields in \ref{bigfoot}); confirmation of this feature is deferred to a future study, as it would require a more advanced analysis, including measurement of radial velocities and matching of our dataset to the Gaia catalog.

Figure \ref{bhbplot} shows how the distribution of BHB stars compares with that of the red giants.  To first order, the BHB appears to more closely follow the spatial distribution of the metal poor $\rm ([Fe/H]<-0.5)$ red clump stars, rather than that of the red giants that concentrate more the plane \citep{bdbs2}. The lower two plots for Fig \ref{bhbplot} were created in the same way as Fig \ref{bhbzoom}, and in fact the middle panel of Fig \ref{bhbplot} is the same as that in Fig \ref{bhbzoom}, but with a different binning.  The upper panel of Fig \ref{bhbplot} simply counts the number of  RGB stars between $16.5<i_0<17.5$, which largely avoids the red clump and also the foreground disk.  The colour map shows the number density within each grid point.  The bottom panel of this figure is simply a straight division of the red giant and
BHB maps that are illustrated in the upper two panels.

\section{Candidate Globular Clusters from the VVV Survey}

The VVV survey combines infrared photometry and variability studies, a powerful combination that potentially revealed previously uncatalogued globular clusters both through enhancements in stellar density and recently, as advanced in \citet{min18} by identifying groups of RR Lyrae stars similar derived distance as representatives of an underlying hosting globular cluster.    This approach to discovering new clusters and systems is indeed interesting, and may have greater power once proper motion surveys extend sufficiently faint that such associations can be made in the kinematic as well as the spatial sense.  However in our case we are not able to confirm any of the new globular clusters inferred from the application of this approach with current data.

We have used our optical BDBS catalog and the Gaia DR2 cross-match to investigate the proposed globular clusters, both from the standpoint of e.g. stellar concentration and BHB stars, and also by searching for concentrations in the Gaia DR2 proper motion vector point diagram.    The use of optical colours offers an additional powerful discriminant compared to \citet{gran19} which also considers whether globular cluster candidates previously reported in the literature are real but uses the VVV photometry with its smaller IR colour baseline.   With the benefit of our optical colours, for example, we show that the colour-magnitude diagram of the cluster Minniti 22 \citep{min18} is not distinct from the metal rich Galactic bulge field population and it is therefore not a metal poor globular cluster.
\begin{figure*}
\centering
\includegraphics[scale=0.3]{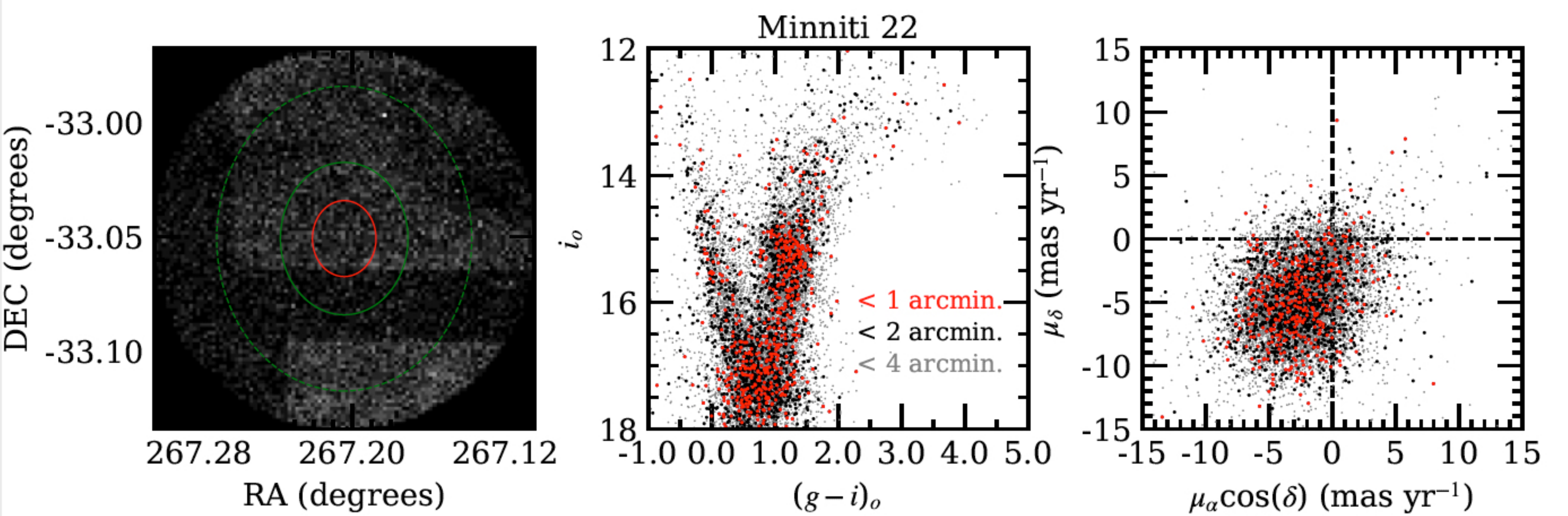}
\caption{Examination of the proposed globular cluster Minniti 22 \citet{min18} identified by  a red giant overdensity and a clustering of RR Lyrae variables; the authors fitted isochrones for a distance and age, and metallicity of ${\rm [Fe/H]}=-1.3$.  The lefthand panel shows the source distribution; the center panel shows the CMD of sources within 1, 2, and 4 arcmin of the cluster center, while the righthand panel shows the vector point plot for those sources.  There is no evidence for a metal poor population, nor for any evidence of a cluster population distinct from the field in any of the 3 plots.  While this confirms the findings of \citet{gran19}, the absence of any distinguishing sequence in the colour-magnitude diagram presents an
additional powerful argument.}
\label{min22}
\end{figure*}

\begin{figure*}
\centering
\includegraphics[scale=0.4]{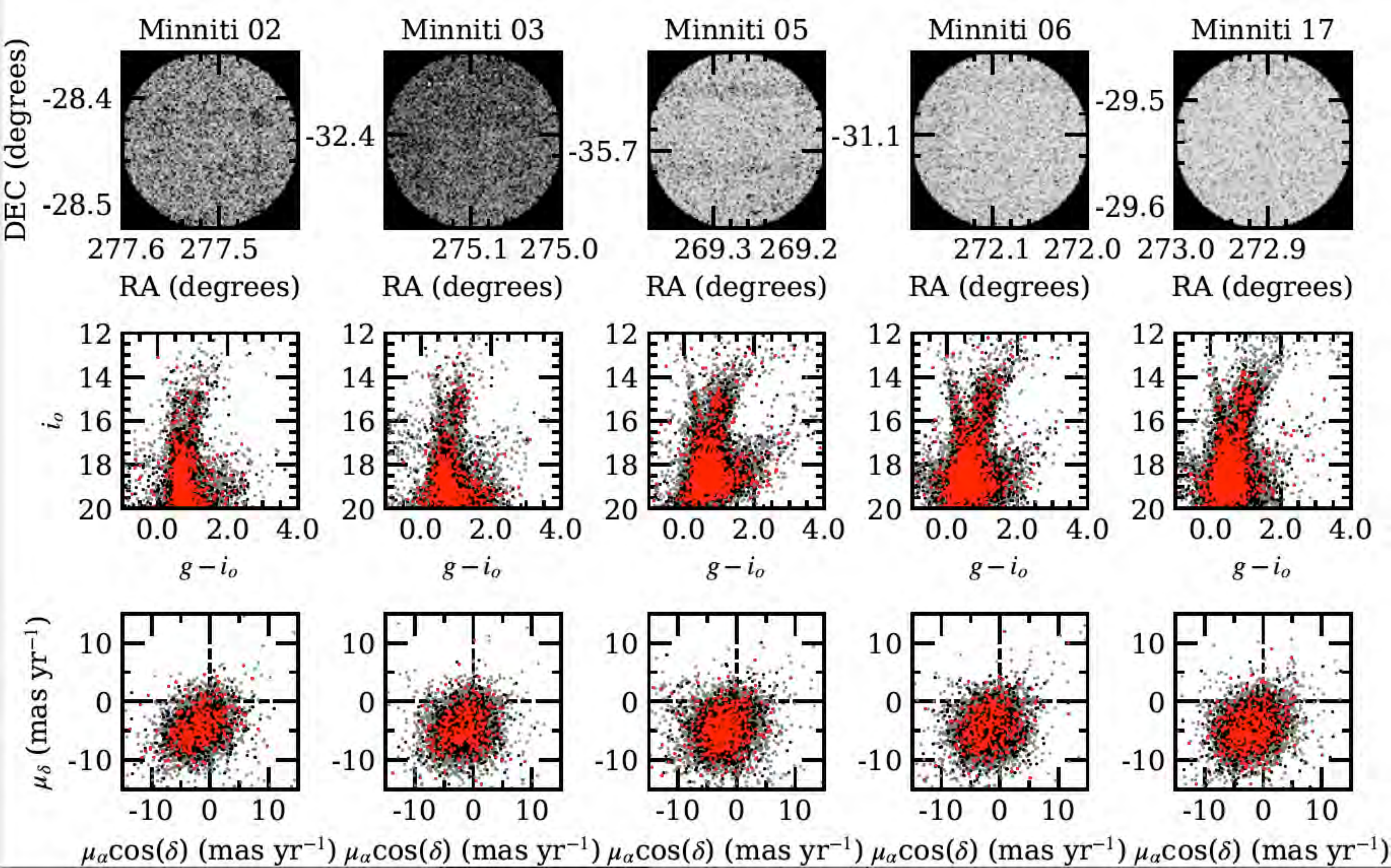}
\caption{The five most luminous cluster candidates from \citet{min17} are illustrated here.  As in Figure \ref{min22}, there is no evidence for the globular clusters in starcounts, colour-magnitude diagrams, or vector point plots.  It is noteworthy that the fields span a wide range of location in the bulge, CMD type, and source density. }
\label{minGCs}
\end{figure*}

Figure \ref{min22} illustrates our investigation of Minniti 22; the authors report a substantial amount of information derived from there analysis of this proposed globular cluster, including ${\rm [Fe/H]}=-1.3$, a distance modulus, and age from isochrone fits.  
Our colour baseline should be sensitive to the metal poor population described in \citet{min18}, but our analysis is unable to confirm it, and also does not isolate a blue horizontal branch.  We show no concentration of red giants or any other sources, and no sign of a cluster in the vector point plot (see \citet{bdbs2} for clear detections of bulge globular clusters in the vector point plot). 

We turn to the 5 most luminous candidates in \citet{min17}, shown in Fig. \ref{minGCs}.    As is the case for Minniti 22, none of these clusters present any evidence for concentration, distinct colour-magnitude diagram, or evidence of a tight cluster in the vector point plot.  We propose that none of these candidates are valid clusters.    We conclude that it is reasonable to ask that any globular cluster candidate in the bulge pass a trifecta of tests- clear concentration, distinct sequences in the CMD, and presence in the vector point plot.    

\section{Conclusions}

We report early results from the Blanco DECam Bulge Survey (BDBS), that spans 200 square degrees with $ugriz$Y photometry.  A companion paper \citep{bdbs2} describes the data reduction methods and calibration, and presents new results on bulge globular clusters and the metallicity distribution derived from red clump stars.  We present a table of globular clusters that lie in the main BDBS footprint.

We present a new array of colour-magnitude diagrams that span the Galactic bulge from $-3^\circ <b < -8^\circ$ and $-8^\circ <l < +8^\circ$.  We detect a bright extension to the red clump that we attribute to the long flat bar; based on Gaia DR2 distances these stars in the foreground relative to the bulk of the bulge.  The signature of the long flat bar is strongest at $(l,b=+8^\circ, -4^\circ)$ but is clearly visible in bulge fields adjacent to this position.
A curious extension of the red clump assigned by \citet{saha19} to massive He burning stars in the bulge instead is due to clump giants 2-4 kpc from the Sun, also likely part of the bar.  The presence of the bar and its complexity will pose a challenge for studies of the main sequence turnoff that aim to
constrain the age of the bulge.  The same features that complicate the red clump will project stars into the regions brighter and bluer than the main sequence turnoff, locations of the colour-magnitude diagram where blue stragglers and trace young/intermediate age populations will reside.

We show new maps of the spatial extent of the bulge blue horizontal branch, including clear detections of a number of known bulge globular clusters, but we find no clear indications of previously unreported streams.  There is a hint of an excess of BHB stars in the extension of the Sagittarius dwarf spheroidal galaxy, but no obvious new streamlike features.  We plan to investigate the possible concentration of BHB stars in the Sgr dwarf vicinity, in future work.

We consider six candidate globular clusters reported in the Galactic bulge, including Minniti 22, "confirmed" as a metal poor globular cluster \citep{min18}.  We find no supporting evidence for these systems based on star counts, colour-magnitude diagrams, and the Gaia DR2 vector point proper motion diagrams.   This strongly confirms the findings of \citet{gran19}, who also failed to confirm $\sim 90$ bulge globular clusters claimed in the literature, but only on the basis of applying Gaia DR2 astrometry to the VVV database.  The combination of BDBS and the optical Gaia DR2 catalog, along with future Gaia data releases, will prove a powerful combination to confirm and challenge claims of new stellar systems toward the bulge.  In the long term, other groundbased datasets such as LSST will strengthen the power to discriminate real from spurious clusters.

With new technology, the prime focus of the CTIO 4-m Blanco telescope is once again, 40 years after its commissioning, helping to launch a new era in the investigation of the Galactic bulge.  We dedicate these efforts to the memory of Victor and Betty Blanco.

\newpage
\clearpage
\section*{Data Availability}
The reduced DECam images on which this article is based are available as given in Table 1, from the data archive of NSF's Optical/IR Laboratory and are found in http://archive1.dm.noao.edu/.
The calibrated photometric catalog data underlying this article will be shared on reasonable request to the corresponding author.   These data may be subject to slight revisions as new versions of our catalog are developed.

\section*{Acknowledgements}
The authors thank an anonymous referee for constructive comments.
RMR, WIC, and CIJ acknowledge partial financial support from NSF grant AST-1411154.
WIC acknowledges support from the Fund for Astrophysical Research, an initiative of the Theodore Dunham, Jr. foundation, and from a grant from the Office of Research and Sponsored Programs, University of Michigan-Dearborn. RMR also acknowledges his late father Jay Baum Rich for financial support.
C. Pilachowski, M. Young, and S. Michaels also acknowledge support from NSF grant AST-1411154.
ITS acknowledges support from the PIFI Grant n. 2018PM0050.
AK gratefully acknowledge funding by the Deutsche Forschungsgemeinschaft (DFG, German Research Foundation) -- Project-ID 138713538 -- SFB 881 
``The Milky Way System'', subproject A11.
AR acknowledges computational support for simulations for the Besancon Galaxy Model, which have been done on the cluster of UTINAM Institute of the Universit\'e de Franche-Comt \'e, supported by the R\'egion de Franche-Comt?e and the french Institut des Sciences de lÕUnivers (INSU).  RMR additionally dedicates this work to the memory of the late Peter Loughrey.

This project used data obtained with the Dark Energy Camera (DECam),
which was constructed by the Dark Energy Survey (DES) collaboration.
Funding for the DES Projects has been provided by 
the U.S. Department of Energy, 
the U.S. National Science Foundation, 
the Ministry of Science and Education of Spain, 
the Science and Technology Facilities Council of the United Kingdom, 
the Higher Education Funding Council for England, 
the National Center for Supercomputing Applications at the University of Illinois at Urbana-Champaign, 
the Kavli Institute of Cosmological Physics at the University of Chicago, 
the Center for Cosmology and Astro-Particle Physics at the Ohio State University, 
the Mitchell Institute for Fundamental Physics and Astronomy at Texas A\&M University, 
Financiadora de Estudos e Projetos, Funda{\c c}{\~a}o Carlos Chagas Filho de Amparo {\`a} Pesquisa do Estado do Rio de Janeiro, 
Conselho Nacional de Desenvolvimento Cient{\'i}fico e Tecnol{\'o}gico and the Minist{\'e}rio da Ci{\^e}ncia, Tecnologia e Inovac{\~a}o, 
the Deutsche Forschungsgemeinschaft, 
and the Collaborating Institutions in the Dark Energy Survey. 
The Collaborating Institutions are 
Argonne National Laboratory, 
the University of California at Santa Cruz, 
the University of Cambridge, 
Centro de Investigaciones En{\'e}rgeticas, Medioambientales y Tecnol{\'o}gicas-Madrid, 
the University of Chicago, 
University College London, 
the DES-Brazil Consortium, 
the University of Edinburgh, 
the Eidgen{\"o}ssische Technische Hoch\-schule (ETH) Z{\"u}rich, 
Fermi National Accelerator Laboratory, 
the University of Illinois at Urbana-Champaign, 
the Institut de Ci{\`e}ncies de l'Espai (IEEC/CSIC), 
the Institut de F{\'i}sica d'Altes Energies, 
Lawrence Berkeley National Laboratory, 
the Ludwig-Maximilians Universit{\"a}t M{\"u}nchen and the associated Excellence Cluster Universe, 
the University of Michigan, 
{the} National Optical Astronomy Observatory, 
the University of Nottingham, 
the Ohio State University, 
the OzDES Membership Consortium
the University of Pennsylvania, 
the University of Portsmouth, 
SLAC National Accelerator Laboratory, 
Stanford University, 
the University of Sussex, 
and Texas A\&M University.

The research presented here is partially supported by the National Key
R\&D Program of China under grant No. 2018YFA0404501; by the National
Natural Science Foundation of China under grant Nos. 11773052,
11761131016, 11333003. J.S. acknowledges support from a {\it Newton Advanced Fellowship} awarded by the Royal Society and the Newton Fund. 
Based on observations at Cerro Tololo Inter-American Observatory (2013A-0529; 2014A-0480; PI:Rich), National Optical
Astronomy Observatory, which is operated by the Association of
Universities for Research in Astronomy (AURA) under a cooperative agreement with the
National Science Foundation.
This work has made use of data from the European Space Agency (ESA) mission
{\it Gaia} (\url{https://www.cosmos.esa.int/gaia}), processed by the {\it Gaia}
Data Processing and Analysis Consortium (DPAC,
\url{https://www.cosmos.esa.int/web/gaia/dpac/consortium}). Funding for the DPAC
has been provided by national institutions, in particular the institutions
participating in the {\it Gaia} Multilateral Agreement.

%%%%%%%%%%%%%%%%%%%% REFERENCES %%%%%%%%%%%%%%%%%%

% The best way to enter references is to use BibTeX:

%\bibliographystyle{mnras}
%\bibliography{example} % if your bibtex file is called example.bib

% Alternatively you could enter them by hand, like this:
% This method is tedious and prone to error if you have lots of references
%\newpage
%\begin{thebibliography}{99}
\bibliographystyle{mnras}
\input{bdbs1.bbl}

%\end{thebibliography}

%%%%%%%%%%%%%%%%%%%%%%%%%%%%%%%%%%%%%%%%%%%%%%%%%%

% Don't change these lines
\bsp	% typesetting comment
\label{lastpage}
\end{document}